\documentclass[twocolumn,showpacs,preprintnumbers,amsmath,amssymb]{revtex4}

\usepackage{graphicx} 
\usepackage{dcolumn} 
\usepackage{bm} 

\usepackage{amssymb} 
\usepackage{amsmath} 

\usepackage{here}

\begin{document}


\title{Evolution of dissipative and non-dissipative universes \\ in holographic cosmological models with a power-law term}

\author{Nobuyoshi {\sc Komatsu}}  \altaffiliation{E-mail: komatsu@se.kanazawa-u.ac.jp} 
\affiliation{Department of Mechanical Systems Engineering, Kanazawa University, Kakuma-machi, Kanazawa, Ishikawa 920-1192, Japan}

\date{\today}

\begin{abstract}

Density perturbations related to structure formations are expected to be different in dissipative and non-dissipative universes, even if the background evolution of the two universes is the same.
To clarify the difference between the two universes, first-order density perturbations are studied, using two types of holographic cosmological models.
The first type is a "$\Lambda(t)$ model" similar to a time-varying $\Lambda(t)$ cosmology for the non-dissipative universe.
The second type is a "BV model" similar to a bulk viscous cosmology for the dissipative universe.  
To systematically examine the two different universes, a power-law term proportional to $H^{\alpha}$ is applied to the $\Lambda(t)$ and BV (bulk-viscous-cosmology-like) models, assuming a flat Friedmann--Robertson--Walker model for the late universe.
Here, $H$ is the Hubble parameter and $\alpha$ is a free parameter whose value is a real number.
The $\Lambda(t)$-$H^{\alpha}$ and BV-$H^{\alpha}$ models are used to examine first-order density perturbations for matter, in which the background evolution of the two models is equivalent.
In addition, thermodynamic constraints on the two models are discussed, with a focus on the maximization of entropy on the horizon of the universe, extending previous analyses [Phys.\ Rev.\ D \textbf{100}, 123545 (2019); \textbf{102}, 063512 (2020)].
Consequently, the $\Lambda(t)$-$H^{\alpha}$ model for small $|\alpha|$ values is found to be consistent with observations and satisfies the thermodynamic constraints, compared with the BV-$H^{\alpha}$ model.
The results show that the non-dissipative universe described by the $\Lambda(t)$-$H^{\alpha}$ model similar to lambda cold dark matter models is likely favored.

\end{abstract}


\pacs{98.80.-k, 95.30.Tg, 98.80.Es}

\maketitle

\section{Introduction}

A paradigm for the cosmic expansion history, that is, an accelerated expansion of the late universe \cite{PERL1998_Riess1998,Suzuki_2012,Hubble2017,Planck2018,Riess20162019}, can be explained by lambda cold dark matter ($\Lambda$CDM) models \cite{Weinberg1,Bamba1}.
The $\Lambda$CDM model assumes an extra driving term related to the cosmological constant $\Lambda$ and an additional energy component called "dark energy."
However, it is well known that the $\Lambda$CDM model suffers from several theoretical difficulties, including the cosmological constant problem and the cosmic coincidence problem \cite{Weinberg1989}.
To resolve these difficulties, various cosmological models have been suggested, such as 
$\Lambda (t)$CDM models [i.e., a time-varying $\Lambda (t)$ cosmology] \cite{Freese,Sola_2009,Nojiri2006,Richarte2013a,Basilakos2018,Sola_2015L14,Valent2015,Sola_2017-2018,Valent2018a,Sola2019,Rezaei2019,Sola_2013b,Sola2020_2,Sola2020_3}, 
bulk viscous models \cite{Weinberg0,Murphy1,Barrow11-Zimdahl1,Brevik1-Nojiri1,Barrow21,Piattella1,Meng3,Avelino2etc,Barbosa2017,Sasidharan,Odintsov2020},
creation of CDM (CCDM) models \cite{Prigogine_1988-1989,Lima1992-2016,Freaza2002,Others2001-2016,Nunes2016B,Barrow20172019,Lima2011,Ramos_2014,Ramos_2014b,Sola2020,Cardenas2020}, 
and thermodynamic scenarios \cite{Easson,Basilakos1,Basilakos2014,Gohar_a_b,Nunes2016,Lepe1,Padma2012AB,Cai2012-Tu2013,Chakraborty-Morad,Neto2018a,Koma457,Koma8,Koma10,Koma11,Koma12,2020Morad,Koma6,Koma9,Koma14,Koma15}.

Formulations of these cosmological models can be categorized into several types, and their theoretical backgrounds are different \cite{Koma6,Koma9,Koma14,Koma15}.
For example, from a dissipative viewpoint, the formulation should be categorized into two types in a Friedmann--Robertson--Walker (FRW) universe.
The first type is $\Lambda(t)$, which is similar to $\Lambda (t)$CDM models \cite{Koma6,Koma9,Koma14,Koma15}.
In $\Lambda(t)$ models, an extra driving term is added to both the Friedmann equation and the Friedmann--Lema\^{i}tre acceleration equation.
The second type is BV, which is similar to both bulk viscous models and CCDM models \cite{Koma6,Koma9,Koma14,Koma15}.
In bulk-viscous-cosmology-like (BV) models, the acceleration equation includes an extra driving term, whereas the Friedmann equation does not.
It is possible to consider that the $\Lambda(t)$ model is related to "reversible entropy," due to, for example, the exchange of matter (energy) \cite{Barrow22,Prigogine_1998}, whereas the BV model is related to "irreversible entropy," due to, for example, gravitationally induced particle creation \cite{Prigogine_1988-1989,Lima1992-2016}.
In this sense, the $\Lambda(t)$ and BV models can describe non-dissipative and dissipative universes, respectively \cite{Koma15}.

The background evolution of the universe in the $\Lambda(t)$ and BV models becomes the same when an equivalent driving term is assumed \cite{Koma15}.
However, even in this case, density perturbations related to structure formations are expected to be different \cite{Koma6,Koma9}.
In fact, the influences of several driving terms, such as constant, Hubble parameter ($H$), and $H^{2}$ terms, have been examined in the $\Lambda(t)$ and BV models.
For the $\Lambda(t)$ model, Basilakos \textit{et al.} reported that the $H^{2}$ terms in $\Lambda(t)$CDM models do not describe structure formations properly \cite{Basilakos1,Sola_2009}.
In contrast, Sol\`{a} \textit{et al.} showed that a combination of the constant and $H^{2}$ terms is favored \cite{Sola_2015L14}.
Such a power series of $H$ has been examined in, for example, the works of G\'{o}mez-Valent \textit{et al.} \cite{Valent2015} and Rezaei \textit{et al.} \cite{Sola2019}.
For the BV model, Li and Barrow have reported that bulk viscous models with the $H$ term are inconsistent with observations of structure formations \cite{Barrow21}.
Barbosa \textit{et al.} used bulk and shear viscous models to point out a similar inconsistency \cite{Barbosa2017}.
In addition, Jesus \textit{et al.} \cite{Lima2011} and Ramos \textit{et al.} \cite{Ramos_2014} showed that CCDM models with a constant term are inconsistent with the observed growth rate for clustering.
(In the CCDM model, a \textit{negative sound speed} \cite{Lima2011} and the existence of \textit{clustered matter} \cite{Ramos_2014} are necessary to properly describe the growth rate \cite{Koma8}.)

In those works, the $\Lambda(t)$ and BV models are separately discussed and, therefore, the dissipative and non-dissipative universes have not yet been examined systematically. 
Accordingly, it is worthwhile to clarify the difference between the dissipative and non-dissipative universes.
To systematically study the two universes, two types of holographic cosmological models that include $H^{\alpha}$ terms \cite{Koma14,Koma15} are suitable.
Here, $\alpha$ is a dimensionless constant whose value is a real number.
The power-law term is obtained from, for example, Padmanabhan's holographic equipartition law \cite{Padma2012AB} with a power-law-corrected entropy \cite{Das2008Radicella2010}, as examined in a previous work \cite{Koma11}.

In this context, we study the evolution of the universe in the $\Lambda(t)$-$H^{\alpha}$ and BV-$H^{\alpha}$ models and discuss observational constraints on the two models.
The universe is expected to behave as an ordinary isolated macroscopic system \cite{Mimoso2013}, where the entropy of the universe does not decrease and approaches a certain maximum value in the last stage \cite{Pavon2013,Krishna20172019,Pavon2019}.
In fact, thermodynamic constraints on the two models have been separately discussed in previous works \cite{Koma14,Koma15}.
Accordingly, we examine the observational constraints in combination with the thermodynamic constraints.
The observational and thermodynamic constraints should provide new insights into a discussion of the dissipative and non-dissipative universes.
Note that this discussion is focused on the late universe and, therefore, the inflation of the early universe is not discussed here.

The remainder of the present paper is organized as follows.
In Sec.\ \ref{Cosmological equations and first-order density perturbations}, cosmological equations and first-order density perturbations for the $\Lambda(t)$ and BV models are reviewed. 
In Sec.\ \ref{Cosmological models with a power-law term}, a power-law term proportional to $H^{\alpha}$ is applied to the $\Lambda(t)$ and BV models.
In Sec.\ \ref{Lambda(t) and BV models with a power-law term}, the $\Lambda(t)$ and BV models that include the $H^{\alpha}$ term, that is, the $\Lambda(t)$-$H^{\alpha}$ and BV-$H^{\alpha}$ models, are formulated.
In Sec.\ \ref{Density perturbations in the present models}, density perturbations in the $\Lambda(t)$-$H^{\alpha}$ and BV-$H^{\alpha}$ models are derived.
Sec.\ \ref{Thermodynamic constraints} reviews thermodynamic constraints on the two models.
In Sec.\ \ref{Evolution of the universe}, the evolution of the universe in the two models is examined.
In addition, the observational and thermodynamic constraints on the two models are investigated.
Finally, Sec.\ \ref{Conclusions} presents the conclusions of the study.

\section{Cosmological equations and first-order density perturbations} 
\label{Cosmological equations and first-order density perturbations}

We consider a homogeneous, isotropic, and spatially flat universe and examine the scale factor $a(t)$ at time $t$ in the FRW metric. 
An expanding universe is assumed from observations \cite{Hubble2017}.

In this section, we review cosmological equations and density perturbations in dissipative and non-dissipative universes, according to previous works \cite{Koma6,Koma9,Koma14,Koma15}.
In Sec.\ \ref{Cosmological equations for L(t) and BV}, cosmological equations for the $\Lambda(t)$ and BV models are presented. 
Sec.\ \ref{Density perturbations for L(t) and BV} reviews first-order density perturbations in the two models.

\subsection{Cosmological equations} 
\label{Cosmological equations for L(t) and BV}

We present formulations of cosmological equations for $\Lambda(t)$ and BV models in a flat FRW universe \cite{Koma6,Koma9,Koma14,Koma15}.
The Friedmann, acceleration, and continuity equations are written as \cite{Koma6,Koma9,Koma14,Koma15} 
\begin{equation}
 H(t)^2      =  \frac{ 8\pi G }{ 3 } \rho (t)    + f_{\Lambda}(t)            ,                                                 
\label{eq:General_FRW01} 
\end{equation} 
\begin{align}
  \frac{ \ddot{a}(t) }{ a(t) }   
                                          &=  -  \frac{ 4\pi G }{ 3 }  ( 1+  3w ) \rho (t)                                   +   f_{\Lambda}(t)    +  h_{\textrm{B}}(t)  , 
\label{eq:General_FRW02}
\end{align}
\begin{equation}
       \dot{\rho} + 3  H (1+w)  \rho   =    -  \frac{3}{8 \pi G }   \dot{f}_{\Lambda}(t)      +    \frac{3 }{4 \pi G}     H h_{\textrm{B}}(t)              , 
\label{eq:drho_General}
\end{equation}
with the Hubble parameter $H(t)$ defined as 
\begin{equation}
   H(t) \equiv   \frac{ da/dt }{a(t)} =   \frac{ \dot{a}(t) } {a(t)}  ,
\label{eq:Hubble}
\end{equation}
where $G$, $\rho(t)$, and $p(t)$ are the gravitational constant, the mass density of cosmological fluids, and the pressure of cosmological fluids, respectively \cite{Koma9}.
Also, $w$ represents the equation of the state parameter for a generic component of matter, which is given as  \cite{Koma14}
\begin{equation}
  w = \frac{ p(t) } { \rho(t)  c^2 }    ,
\label{eq:w}
\end{equation}
where $c$ represents the speed of light.
For a matter-dominated universe and a radiation-dominated universe, the values of $w$ are $0$ and $1/3$, respectively.
Here, we consider a matter-dominated universe, that is, $w =0$, and
we neglect the influence of radiation in the late universe.

In the above formulation, two extra driving terms, $f_{\Lambda}(t)$ and $h_{\textrm{B}}(t)$, are phenomenologically assumed \cite{Koma14}.
Specifically, $f_{\Lambda}(t)$ is used for $\Lambda (t)$ models and $h_{\textrm{B}}(t)$ is used for BV models.
Accordingly, we set $h_{\textrm{B}}(t) =0$ for the $\Lambda (t)$ model and $f_{\Lambda}(t) = 0$ for the BV model \cite{Koma14,Koma15}.

In a matter-dominated universe ($w=0$), coupling Eq.\ (\ref{eq:General_FRW01}) with Eq.\ (\ref{eq:General_FRW02}) yields \cite{Koma14,Koma15} 
\begin{equation}
    \dot{H} = - \frac{3}{2}  H^{2}  +  \frac{3}{2}    f_{\Lambda}(t)     + h_{\textrm{B}}(t)   ,
\label{eq:Back2}
\end{equation}
or equivalently, 
\begin{equation}
    \dot{H} =
    \begin{cases} 
          - \frac{3}{2}  H^{2}  +  \frac{3}{2}    f_{\Lambda}(t)                        &   (\Lambda(t)    \hspace{1mm} \rm{model})  ,         \\
          - \frac{3}{2}  H^{2}  +                     h_{\textrm{B}}(t)                     &   (\rm{BV} \hspace{1mm} \rm{model})           .         \\
    \end{cases}
\label{eq:Back2_2}
\end{equation}
These equations indicate that the background evolution of the universe in the $\Lambda(t)$ and BV models is equivalent when the driving terms are equal \cite{Koma15}:
\begin{equation}
    \frac{3}{2}  f_{\Lambda}(t) = h_{\textrm{B}}(t)     . 
\label{eq:32f=h}
\end{equation}
In this study, we set $\frac{3}{2}  f_{\Lambda}(t) = h_{\textrm{B}}(t)$, as shown in Eq.\ (\ref{eq:32f=h}), and therefore, the background evolution of the universe in the two models is the same.
In addition, a power-law term proportional to $H^{\alpha}$ is used for the driving term.
The power-law term is discussed in Sec.\ \ref{Cosmological models with a power-law term}.

\subsection{First-order density perturbations}
\label{Density perturbations for L(t) and BV}

In this subsection, we present two formulations to examine density perturbations in the $\Lambda(t)$ and BV models.
In Secs.\ \ref{Formulations for the Lambda(t) model} and \ref{Formulations for the BV model}, we review density perturbations in the $\Lambda(t)$ and BV models, respectively.
The formulation for the $\Lambda(t)$ and BV models used in this study is based on the works of Basilakos \textit{et al.} \cite{Sola_2009} and Jesus \textit{et al.} \cite{Lima2011}, respectively.
The two formulations were summarized in a previous report \cite{Koma6}, using the neo-Newtonian approach proposed by Lima \textit{et al.} \cite{Lima_Newtonian_1997}.
A unified formulation based on the neo-Newtonian approach is summarized in Appendix\ \ref{Unified formulation}.

In this paper, we examine first-order density perturbations in the linear approximation by assuming a matter-dominated universe ($w=0$), which is equivalent to a fluid without pressure ($p=0$) \cite{Koma6}. 
In other words, we focus on density perturbations for matter and neglect other perturbations, as discussed below.

\subsubsection{Formulations for the $\Lambda(t)$ model}
\label{Formulations for the Lambda(t) model}

Density perturbations in $\Lambda(t)$CDM models have been examined in various studies, including Basilakos \textit{et al.} \cite{Sola_2009},  Sol\`{a} \textit{et al.} \cite{Sola_2015L14}, G\'{o}mez-Valent \textit{et al.} \cite{Valent2015}, and Rezaei \textit{et al.} \cite{Sola2019}.
The formulation for the $\Lambda(t)$ model discussed here is essentially equivalent to that for the $\Lambda(t)$CDM model, although the theoretical backgrounds are different.
Therefore, we review density perturbations in the $\Lambda(t)$ model, according to Ref.\ \cite{Sola_2009} and a previous work \cite{Koma6}.

In a matter-dominated universe, substituting $w=0$ into Eq.\ (\ref{eq:drho_General}) yields
\begin{equation}
      \dot{\rho} + 3  H   \rho   =  - \frac{3}{8 \pi G}  \dot{f}_{\Lambda} (t)   ,
\label{eq:drho_L(t)}
\end{equation}
where ${f}_{\Lambda} (t)$ is a general driving term and $h_{\textrm{B}} (t) =0$ is used for the $\Lambda(t)$ model. 
The right-hand side of Eq.\ (\ref{eq:drho_L(t)}) is zero when ${f}_{\Lambda} (t)$ is constant.
(A power-law term is discussed in Sec.\ \ref{Cosmological models with a power-law term}.)

In Ref.\ \cite{Sola_2009}, Basilakos \textit{et al.} focused on models in which the time dependence of $\Lambda(t)$ appears always at the expense of an interaction with matter \cite{Koma6}.
The model is considered to be an energy exchange cosmology that assumes the transfer of energy (matter) between two fluids \cite{Barrow22}.
Similarly, in holographic cosmological models, we assume an interchange of energy between the bulk (the universe) and the boundary (the horizon of the universe) \cite{Lepe1}, as if it is an energy exchange cosmology \cite{Koma6}.
(For example, cosmological equations can be derived from the expansion of cosmic space due to the difference between the degrees of freedom on the boundary and in the bulk by applying the holographic equipartition law with an associated entropy on the horizon \cite{Padma2012AB}.
Interacting holographic dark energy models were examined in Ref.\ \cite{Richarte2013a}.)

Consequently, the time evolution equation for the matter density contrast $\delta \equiv \delta \rho_{m} /\rho_{m} $, namely the perturbation growth factor, is given by \cite{Waga1994}
\begin{equation}
      \ddot{\delta} +  \left ( 2 H + Q \right ) \dot{\delta} - \left [ 4 \pi G \rho  - 2 H Q -\dot{Q} \right ] \delta = 0 ,
\label{eq:density_L(t)}
\end{equation}
where 
\begin{equation}
 \rho =   \frac{3}{8 \pi G}  [H^{2} - {f}_{\Lambda} (t) ],     \quad   Q =  - \frac{3}{8 \pi G}  \frac{ \dot{f}_{\Lambda} (t) }{  \rho  }  ,
\label{eq:Q_L(t)}
\end{equation}
and $\rho$ is the mass density of matter. 
Specifically, $\rho_{m}$ is replaced by $\rho$ because a matter-dominated universe is considered \cite{Koma6}.
In addition, $\rho$ in Eq.\ (\ref{eq:density_L(t)}) represents $\bar{\rho}$, corresponding to a homogenous and isotropic solution for the unperturbed equations \cite{Koma6}.
Substituting Eq.\ (\ref{eq:Q_L(t)}) into Eq.\ (\ref{eq:drho_L(t)}) yields
\begin{equation}
      \dot{\rho} + 3  H  \rho   =   Q \rho  . 
\label{eq:drho_L(t)_Q}
\end{equation}

In this study, for numerical purposes, we use an independent variable \cite{Lima2011} that is defined as 
\begin{equation}
\eta \equiv \ln [\tilde{a}(t)] ,
\label{eta_def}
\end{equation}
where $\tilde{a}(t)$ is the normalized scale factor given by 
\begin{equation}
   \tilde{a} = \frac{a} { a_{0}}      
\label{eq:a_a0}
\end{equation}
and $a_{0}$ is the scale factor at the present time \cite{Koma15}.
From this definition, $\dot{\delta}$ and $\ddot{\delta}$ are written as
\begin{equation}
     \dot{\delta} = H \frac{d \delta}{d \eta} = H \delta^{\prime}  \quad \textrm{and} \quad  \ddot{\delta} =  H^{2} \delta^{\prime \prime}  + H^{\prime} H \delta^{\prime}   ,
\label{delta12}
\end{equation}
where $^{\prime}$ represents the differential with respect to $\eta$, namely $d/d \eta$.
Equation\ (\ref{delta12}) can be written as $\dot{x}  = H x^{\prime}$ and  $ \ddot{x} =  H^{2} x^{\prime \prime}  + H^{\prime} H x^{\prime}$ by using an arbitrary variable $x$.
Applying these equations to Eq.\ (\ref{eq:density_L(t)}) and performing several calculations, the differential equation is written as
\begin{equation}
\delta^{\prime \prime}  + F_{\Lambda}(\eta) \delta^{\prime}  +  G_{\Lambda} (\eta) \delta =0, 
\label{eq:delta-eta_L(t)_0}
\end{equation}
where $F_{\Lambda} (\eta)$ and $G_{\Lambda} (\eta)$ are given by
\begin{equation}
F_{\Lambda} (\eta) =  2  +  \frac{ Q + H^{\prime} }{ H }   
\label{eq:FL(eta)_0}
\end{equation}
and
\begin{align}
G_{\Lambda} (\eta)  =  &   - \frac{3}{2}  + \frac{ 2 Q + Q^{\prime} }{ H } + \frac{  3 f_{\Lambda} (t) }{ 2 H^{2} }     ,
\label{eq:GL(eta)_0}
\end{align}
where $H \neq 0$ and $4 \pi G \rho = \frac{3}{2}( H^{2} - {f}_{\Lambda} (t) )$, as given by Eq.\ (\ref{eq:General_FRW01}).
We use Eqs.\ (\ref{eq:delta-eta_L(t)_0})--(\ref{eq:GL(eta)_0}) to examine density perturbations in the $\Lambda (t)$ model.

It should be noted that it is necessary to define explicitly the functional form of the ${f}_{\Lambda}  (t)$ component to solve the above differential equation \cite{Koma6}.
As described in Ref.\ \cite{Sola_2009}, the approach based on Eq.\ (\ref{eq:density_L(t)}) implies that dark energy perturbations are assumed to be negligible. 
This assumption is generally justified in most cases \cite{Sola_2009,Sola_2007-2009} and
has been most recently examined in the work of G\'{o}mez-Valent and Sol\`{a} \cite{Valent2018a}.
In this study, we assume that boundary perturbations are negligible in holographic cosmological models \cite{Koma6}.

\subsubsection{Formulations for the BV model}
\label{Formulations for the BV model}

BV models assume dissipation processes and therefore, the formulation for the BV model is essentially equivalent to that for both bulk viscous and CCDM models \cite{Koma6,Koma9,Koma14,Koma15}.
For example, density perturbations in the CCDM model were examined by Jesus \textit{et al.} \cite{Lima2011}.
In addition, density perturbations in the BV model can be derived from a neo-Newtonian approach \cite{Koma6}.
The neo-Newtonian approach was proposed by Lima \textit{et al.} \cite{Lima_Newtonian_1997}, following earlier ideas developed by McCrea \cite{McCrea_1951} and Harrison \cite{Harrison_1965}, to describe a Newtonian universe with pressure \cite{Lima2011}.
We apply the ideas in these works to review the density perturbations in the BV model.

We assume a matter-dominated universe ($w=0$) and model it as a pressureless fluid ($p=0$). 
Substituting $w=0$ into Eq.\ (\ref{eq:drho_General}) yields  \cite{Koma6}
\begin{equation}
      \dot{\rho} + 3  H   \rho   =   \frac{3}{4 \pi G}  H h_{\textrm{B}} (t)   ,
\label{eq:drho_General_BV}
\end{equation}
where $ h_{\textrm{B}} (t)$ is a general driving term and ${f}_{\Lambda} (t) =0$ is used for the BV model.
(A power-law term is discussed in Sec.\ \ref{Cosmological models with a power-law term}.)
The right-hand side of Eq.\ (\ref{eq:drho_General_BV}) is not $0$, even if $ h_{\textrm{B}} (t)$ is constant.
Equation\ (\ref{eq:drho_General_BV}) is essentially equivalent to the model examined in Ref.\ \cite{Lima2011}.
To confirm this, we consider an effective pressure $p_{e}$ due to dissipation processes \cite{Koma6}.
The effective pressure $p_{e}$ is given by $p_{e}   =  p + p_{c} $, where $p_{c}$ is the creation pressure for constant specific entropy in CCDM models \cite{Lima2011}.
In this study, $p_{e}$ is equivalent to $p_{c}$, because $p=0$.
In addition, as examined in Ref.\ \cite{Lima2011}, $p_{c}$ for a CDM component can be written as 
\begin{equation}
p_{c} = - \frac{\rho c^2 \Gamma}{3H}    , 
\label{eq:pc_BV}
\end{equation}
where $\Gamma$ is given by
\begin{equation}
\Gamma = \frac{3}{4 \pi G}  \frac{H  h_{\textrm{B}} (t)}{ \rho }      .
\label{eq:gamma_BV}
\end{equation}
Therefore, Eq.\ (\ref{eq:drho_General_BV}) is written as 
\begin{equation}
      \dot{\rho} + 3 H \rho   =   \Gamma \rho  ,
\label{eq:drho_General_BV_gamma}
\end{equation}
where $\Gamma$ is a parameter related to entropy production processes \cite{Koma15}.
In the CCDM model, $\Gamma$ is considered to be the creation rate of CDM particles \cite{Lima2011}.

We note that a perturbation analysis in cosmology generally requires a full relativistic description, as examined in Jesus \textit{et al.} \cite{Lima2011}.
This is because the standard non-relativistic (Newtonian) approach works well only when the scale of perturbation is much less than the Hubble radius and the velocity of peculiar motions is small in comparison with the Hubble flow \cite{Lima2011}. 
In fact, Jesus \textit{et al.} proposed a neo-Newtonian approximation that circumvents such difficulties.
In this study, we applied a neo-Newtonian approximation to the BV model, as discussed in Ref.\ \cite{Koma6}.

In our units, $c=1$ and the time evolution equation for the matter density contrast $\delta$ is given by \cite{Lima2011} 
\begin{align}
\ddot{\delta}  & + \left [ 2 H + \Gamma + 3 c_{\rm{eff}}^{2} H  - \frac{ \Gamma \dot{H} - H \dot{\Gamma} }{ H (3H -\Gamma)}  \right ] \dot{\delta}        \notag \\
                     & +  \Bigg \{      3 (\dot{H} + 2 H^{2}) \left (   c_{\rm{eff}}^{2}  + \frac{ \Gamma }{ 3H } \right )     \notag \\
                     & + 3 H  \left [ \dot{c}_{\rm{eff}}^{2}   - ( 1 + c_{\rm{eff}}^{2}) \frac{ \Gamma \dot{H} - H \dot{\Gamma} }{ H (3H -\Gamma)}  \right ] \notag \\     
                     &  - 4 \pi G \rho \left ( 1 -  \frac{ \Gamma }{ 3H } \right ) ( 1 + 3 c_{\rm{eff}}^{2} )  + \frac{  k^{2} c_{\rm{eff}}^{2} }{ a^{2} }    \Bigg \}     \delta = 0              .
\label{eq:delta-t_BV}
\end{align}
In this study, the effective sound speed, $c_{\rm{eff}}^{2} \equiv  \delta p_{c} /\delta \rho $, is set to 
\begin{equation}
 c_{\rm{eff}}^{2}  \equiv  \frac{\delta p_{c} }{\delta \rho } =0    
\label{eq:ceff2_BV_0}
\end{equation}
to ensure equivalence between the neo-Newtonian and general relativistic approaches \cite{Reis_2003}.
In fact, the neo-Newtonian equation given by Eq.\ (\ref{eq:delta-t_BV}) is equivalent to the general relativistic equation for a single-fluid-dominated universe only when $c_{\rm{eff}}^{2} = 0$, as examined by Reis \cite{Reis_2003}. 
The equivalence is discussed by Ramos \textit{et al.} \cite{Ramos_2014}.

For numerical purposes, we use an independent variable $\eta \equiv \ln [\tilde{a}(t)]$, which is defined by Eq.\ (\ref{eta_def}).
Therefore, $\dot{\delta}$ and $\ddot{\delta}$ are given by Eq.\ (\ref{delta12}).
Also, the Friedmann equation for the BV model is written as $4 \pi G \rho = 3 H^{2}/2 $.
We apply these equations and $c_{\rm{eff}}^{2} = 0$ to Eq.\ (\ref{eq:delta-t_BV}) and perform  several operations to obtain  
\begin{equation}
\delta^{\prime \prime}  + F_{\textrm{B}} (\eta) \delta^{\prime}  +  G_{\textrm{B}} (\eta) \delta =0, 
\label{eq:delta-eta_c=0_BV_1}
\end{equation}
where $F_{\textrm{B}} (\eta)$ and $G_{\textrm{B}} (\eta)$ are given by \cite{Lima2011} 
\begin{equation}
F_{\textrm{B}} (\eta) =  2  + \frac{ \Gamma + H^{\prime} }{ H }   -   \frac{  \Gamma H^{\prime} - H \Gamma^{\prime}  }{ H (3H -\Gamma) }   ,
\label{eq:FB(eta)_c=0_1}
\end{equation}
\begin{equation}
G_{\textrm{B}} (\eta)  =  \left (  \frac{ \Gamma }{ H }  - 1   \right )   \left (  \frac{ \Gamma }{ 2H }  + \frac{3}{2}   \right )   -   \frac{ 3 ( \Gamma H^{\prime} - H \Gamma^{\prime} ) }{ H (3H -\Gamma) }   .
\label{eq:GB(eta)_c=0_1}
\end{equation}
Equations\ (\ref{eq:delta-eta_c=0_BV_1})--(\ref{eq:GB(eta)_c=0_1}) are used to examine density perturbations in the BV model.

In this section, we reviewed density perturbations in the $\Lambda(t)$ and BV models without using a power-law term proportional to $H^{\alpha}$.
In the next section, we apply the power-law term and derive density perturbations in the $\Lambda(t)$-$H^{\alpha}$ and BV-$H^{\alpha}$ models.

\section{$\Lambda(t)$-$H^{\alpha}$ and BV-$H^{\alpha}$ models} 
\label{Cosmological models with a power-law term}

This section discusses two types of cosmological models with a power-law term proportional to $H^{\alpha}$.
In Sec.\ \ref{Lambda(t) and BV models with a power-law term}, the $\Lambda(t)$-$H^{\alpha}$ and BV-$H^{\alpha}$ models are formulated.
In Sec.\ \ref{Density perturbations in the present models}, density perturbations in the two models are derived.
Finally, Sec.\ \ref{Thermodynamic constraints} uses Refs.\ \cite{Koma14,Koma15} to review thermodynamic constraints on the two models.

\subsection{Cosmological equations for the $\Lambda(t)$-$H^{\alpha}$ and BV-$H^{\alpha}$ models} 
\label{Lambda(t) and BV models with a power-law term}

General formulations of cosmological equations for the $\Lambda(t)$ and BV models are described in the previous section.
In this section, a power-law term proportional to $H^{\alpha}$ is applied to the $\Lambda(t)$ and BV models.

In fact, cosmological equations can be derived from the expansion of cosmic space due to the difference between the degrees of freedom on the surface and in the bulk, using Padmanabhan's holographic equipartition law with an associated entropy on the horizon \cite{Padma2012AB}.
As examined in a previous work \cite{Koma11}, an acceleration equation that includes $H^{\alpha}$ terms is derived using the holographic equipartition law with entropy corrected by a power law \cite{Das2008Radicella2010}.
For the derivation, see Ref.\ \cite{Koma11}.

The power-law term, namely the $H^{\alpha}$ term, was investigated for a non-dissipative universe based on $\Lambda (t)$ models \cite{Koma11,Koma14} and a dissipative universe based on BV models \cite{Koma15}.
In this paper, the power-law term is applied to the $\Lambda(t)$ and BV models, which
we call the $\Lambda (t)$-$H^{\alpha}$ model and the BV-$H^{\alpha}$ model, respectively.

For the $\Lambda (t)$-$H^{\alpha}$ model, a driving term $f_{\Lambda}(t)$ is given by
\begin{equation}
       f_{\Lambda}(t)        =  \Psi_{\alpha} H_{0}^{2} \left (  \frac{H}{H_{0}} \right )^{\alpha}  , 
\label{eq:fhB(L)}
\end{equation}
and for the BV-$H^{\alpha}$ model, the driving term $h_{\textrm{B}}(t)$ is 
\begin{equation}
       h_{\textrm{B}}(t)     =   \frac{3}{2}  \Psi_{\alpha} H_{0}^{2} \left (  \frac{H}{H_{0}} \right )^{\alpha}    ,
\label{eq:fhB(BV)}
\end{equation}
where $H_{0}$ represents the Hubble parameter at the present time and $\alpha$ is a dimensionless constant whose value is a real number \cite{Koma14,Koma15}.
Also, $\Psi_{\alpha}$ is a density parameter for effective dark energy and is assumed to be 
\begin{equation}
       0 \leq \Psi_{\alpha} \leq 1 .
\label{eq:Psi_01}
\end{equation}
In this paper, $\alpha$ and $\Psi_{\alpha}$ are considered to be independent free parameters  \cite{Koma14,Koma15}.
This means that we phenomenologically assume the power-law term for the two models.
In addition, as shown in Eqs.\ (\ref{eq:fhB(L)}) and (\ref{eq:fhB(BV)}), the two driving terms are set to $\frac{3}{2}  f_{\Lambda}(t) = h_{\textrm{B}}(t)$ so that Eq.\ (\ref{eq:32f=h}) can be satisfied.
Substituting Eq.\ (\ref{eq:fhB(L)}) for the $\Lambda (t)$-$H^{\alpha}$ model into Eq.\ (\ref{eq:Back2_2}) yields
\begin{align}
    \dot{H}    &= - \frac{3}{2} H^{2}  \left [  1 -   \Psi_{\alpha} \left (  \frac{H}{H_{0}} \right )^{\alpha -2} \right ]      . 
\label{eq:Back_power_fLhB}
\end{align} %
An equivalent equation is obtained by substituting Eq.\ (\ref{eq:fhB(BV)}) for the BV-$H^{\alpha}$ model into Eq.\ (\ref{eq:Back2_2}).
In this way, the same background evolution of the universe is established for both the $\Lambda (t)$-$H^{\alpha}$ and BV-$H^{\alpha}$ models.

Equation\ (\ref{eq:Back_power_fLhB}) has been examined previously \cite{Koma14,Koma15}.
The solution for $\alpha \neq 2$ is written as 
\begin{equation}  
    \left ( \frac{H}{H_{0}} \right )^{2-\alpha}  =   (1- \Psi_{\alpha})   \tilde{a}^{ - \frac{3(2-\alpha)}{2}  }  + \Psi_{\alpha}   ,
\label{eq:Sol_HH0_power}
\end{equation}
and the solution for $\alpha = 2$ is
\begin{equation}
     \frac{H}{H_{0}}  =     \tilde{a}^{ - \frac{3 (1- \Psi_{\alpha}) }{2}  }     ,
\label{eq:Sol_HH0_aa0_H2}
\end{equation}
where $\tilde{a}$ is the normalized scale factor given by Eq.\ (\ref{eq:a_a0}).
These solutions can be applied to the $\Lambda (t)$-$H^{\alpha}$ and BV-$H^{\alpha}$ models.
The derivation is summarized in Ref.\ \cite{Koma14}. (When $\alpha =0$, replacing $\Psi_{\alpha}$ by $\Omega_{\Lambda}$, the density parameter for $\Lambda$, gives a background evolution that is equivalent to that in $\Lambda$CDM models \cite{Koma15}. 
We neglect the influence of radiation in a late, flat FRW universe.)
In addition, the temporal deceleration parameter $q$, defined by $q \equiv  - \left ( \frac{\ddot{a} } {a H^{2}} \right ) $, can be calculated from the above equations.
Using the result of Refs.\ \cite{Koma14,Koma15}, the deceleration parameter $q$ for the two models is written as 
\begin{align}
   q    &= \frac{1}{2}  -   \frac{3}{2}     \Psi_{\alpha} \left [  (1- \Psi_{\alpha})   \tilde{a}^{ - \frac{3(2-\alpha)}{2}  }  + \Psi_{\alpha}    \right ]^{-1}    ,
\label{eq:q_power}
\end{align}
where a positive and negative $q$ represent deceleration and acceleration, respectively.

In the present study, the background evolution of the universe in the $\Lambda (t)$-$H^{\alpha}$ and BV-$H^{\alpha}$ models is set to be the same, as mentioned previously.
However, even in this case, the evolution of density perturbations is expected to be different.
We discuss density perturbations in the next subsection.
Hereafter, we consider $\alpha \neq 2$ because the result for $\alpha \neq 2$ reduces to that for $\alpha = 2$ when $\alpha \rightarrow 2$.

It should be noted that a power series of $H$ for $\Lambda(t)$ models was examined in works such as Sol\`{a} \textit{et al.} \cite{Sola_2015L14}, G\'{o}mez-Valent \textit{et al.} \cite{Valent2015}, and Rezaei \textit{et al.} \cite{Sola2019}.
For BV models, a power-law term was examined in, for example, the works of Freaza \textit{et al.} \cite{Freaza2002}, Ramos \textit{et al.} \cite{Ramos_2014}, and C\'{a}rdenas \textit{et al.} \cite{Cardenas2020}. 
From a microscopic viewpoint, the driving term for running vacuum models [related to $\Lambda(t)$ models] can be obtained from various concepts, including renormalization group equations \cite{Sola_2013b}, quantum field theory in curved spacetime \cite{Sola2020_2}, and string theory \cite{Sola2020_3}.
In contrast, the driving term for bulk viscous and CCDM models (related to BV models) is phenomenologically assumed based on macroscopic properties, such as bulk viscosity of cosmological fluids \cite{Weinberg0,Murphy1} and matter creation \cite{Prigogine_1988-1989,Lima1992-2016}, respectively.
In this paper, we phenomenologically assume the power-law term for the $\Lambda (t)$ and BV models and therefore, the theoretical backgrounds are different from those of the above models.
However, the formulations used here are essentially equivalent to those for the models.

\subsection{First-order density perturbations in the $\Lambda(t)$-$H^{\alpha}$ and BV-$H^{\alpha}$ models} 
\label{Density perturbations in the present models}

In this section, we examine first-order density perturbations in the $\Lambda(t)$-$H^{\alpha}$ and BV-$H^{\alpha}$ models.
To calculate density perturbations, $H^{\prime}/H$ is required.
Accordingly, we write the background evolution again.
From Eq.\ (\ref{eq:Sol_HH0_power}), the evolution of the Hubble parameter for $\alpha \neq 2$ can be written as 
\begin{align}  
     \frac{H}{H_{0}}   &=  \left [  (1- \Psi_{\alpha})   \tilde{a}^{ - \frac{3(2-\alpha)}{2}  }  + \Psi_{\alpha}   \right ]^ { \frac{1}{2-\alpha} }  \notag \\
                            &=  \left [  (1- \Psi_{\alpha})   \tilde{a}^{ - \beta  }      + \Psi_{\alpha}   \right ]^ { \frac{1}{2-\alpha} }                    \notag \\   
                            &=  \left [  (1- \Psi_{\alpha})            e^{- \beta \eta}   + \Psi_{\alpha}   \right ]^ { \frac{1}{2-\alpha} }                 ,       
\label{eq:Sol_HH0_power_3}
\end{align}
where $\tilde{a}^{- \beta}$ is replaced by $e^{- \beta \eta}$ using $\eta \equiv \ln [\tilde{a}(t)]$ given by Eq.\ (\ref{eta_def}).
A dimensionless parameter $\beta$ is used for simplicity and given by
\begin{equation}  
                                 \beta  = \frac{3(2-\alpha)}{2} .
\label{eq:beta_0}
\end{equation}
Differentiating Eq.\ (\ref{eq:Sol_HH0_power_3}) with respect to $\eta$ yields
\begin{align}  
\frac{ H^{\prime} }{H_{0}} &= \frac{d}{d \eta} \left ( \frac{H}{H_{0}} \right )  =  \frac{d}{d \eta} \left [  (1- \Psi_{\alpha})          e^{- \beta \eta}   + \Psi_{\alpha}   \right ]^ { \frac{1}{2-\alpha} }                                            \notag \\
                                    &=  \frac{(- \beta) (1-  \Psi_{\alpha}) e^{- \beta \eta}   }{2-\alpha}    \left [  (1- \Psi_{\alpha})          e^{- \beta \eta}   + \Psi_{\alpha}   \right ]^ { \frac{1}{2-\alpha} -1 }                                         \notag \\
                                    &=  \frac{- 3 (1-  \Psi_{\alpha}) e^{- \beta \eta} }{2}                         \left [  (1- \Psi_{\alpha})          e^{- \beta \eta}   + \Psi_{\alpha}   \right ]^ { \frac{\alpha -1}{2-\alpha}  }                   .
\label{eq:Hp-H0_power}
\end{align}
Dividing Eq.\ (\ref{eq:Hp-H0_power}) by Eq.\ (\ref{eq:Sol_HH0_power_3}) yields
\begin{align}  
\frac{H^{\prime} }{H}          &= \frac{ \frac{- 3 (1-  \Psi_{\alpha}) e^{- \beta \eta} }{2}                         \left [  (1- \Psi_{\alpha})          e^{- \beta \eta}   + \Psi_{\alpha}   \right ]^ { \frac{\alpha -1}{2-\alpha}  }       }
                                                     {  \left [  (1- \Psi_{\alpha})          e^{- \beta \eta}   + \Psi_{\alpha}   \right ]^ { \frac{1}{2-\alpha}  }                                                                                                               }  \notag \\ 
                                         &= \frac{ -\frac{3}{2} (1- \Psi_{\alpha})    e^{- \beta \eta}       }{   (1-  \Psi_{\alpha})e^{- \beta \eta}       +  \Psi_{\alpha}                                  }                                                            
                                         = \frac{ -\frac{3}{2} ( 1- \Psi_{\alpha} )                               }{   (1-  \Psi_{\alpha})                              +  \Psi_{\alpha} e^{\beta \eta}            }    .
\label{eq:Hp-H_power_a_eta_1}
\end{align}
Then, we use the obtained $H^{\prime} / H$ to derive density perturbations in the $\Lambda(t)$-$H^{\alpha}$ and BV-$H^{\alpha}$ models.
Note that $\alpha \neq 2$ is considered here because the result for $\alpha \neq 2$ reduces to that for $\alpha = 2$ when $\alpha \rightarrow 2$.

\subsubsection{Density perturbations in the $\Lambda(t)$-$H^{\alpha}$ model} 
\label{DensityL(t)HA}

We examine density perturbations in the $\Lambda(t)$-$H^{\alpha}$ model.
From Eq.\ (\ref{eq:fhB(L)}), the driving term $f_{\Lambda}(t)$ is written as 
\begin{equation}
       f_{\Lambda}(t)    =  \Psi_{\alpha} H_{0}^{2} \left (  \frac{H}{H_{0}} \right )^{\alpha}  .  
\label{eq:fhB(L)_2}
\end{equation}
We write Eqs.\ (\ref{eq:delta-eta_L(t)_0})--(\ref{eq:GL(eta)_0}) again as
\begin{equation}
\delta^{\prime \prime}  + F_{\Lambda}(\eta) \delta^{\prime}  +  G_{\Lambda} (\eta) \delta =0, 
\label{eq:delta-eta_L(t)_1}
\end{equation}
\begin{equation}
F_{\Lambda} (\eta) =  2  +  \frac{ Q + H^{\prime} }{ H }   ,
\label{eq:FL(eta)_1}
\end{equation}
\begin{align}
G_{\Lambda} (\eta)  =  &   - \frac{3}{2}  + \frac{ 2 Q + Q^{\prime} }{ H } + \frac{  3 f_{\Lambda} (t) }{ 2 H^{2} }     ,
\label{eq:GL(eta)_1}
\end{align}
where $Q$ and $\rho$ given by Eq.\ (\ref{eq:Q_L(t)}) are written as
\begin{equation}
 Q =  - \frac{3}{8 \pi G}  \frac{ \dot{f}_{\Lambda} (t) }{  \rho  }  ,
\label{eq:Q_L(t)_2}
\end{equation}
\begin{equation}
 \rho =   \frac{3}{8 \pi G}  [H^{2} - {f}_{\Lambda} (t) ]. 
\label{eq:rho_L(t)}
\end{equation}

We now calculate $F_{\Lambda} (\eta)$ and $G_{\Lambda} (\eta)$.
To this end, we require three terms, namely $Q / H$, $Q^{\prime} / H$, and $f_{\Lambda} (t) / H^{2} $.
First, we calculate $Q$ to obtain $Q/H$.
Substituting Eq.\ (\ref{eq:rho_L(t)}) into Eq.\ (\ref{eq:Q_L(t)_2}) and applying Eq.\ (\ref{eq:fhB(L)_2}) yields
\begin{align}
    Q  &=  - \frac{3}{8 \pi G}  \frac{ \dot{f}_{\Lambda} (t) }{  \rho  }  = - \frac{3}{8 \pi G}  \frac{ \dot{f}_{\Lambda} (t) }{  \frac{3}{8 \pi G}  (H^{2} - {f}_{\Lambda} (t) ) }                \notag \\
        &=  - \frac{ \alpha \Psi_{\alpha} H_{0} \left (  \frac{H}{H_{0}} \right )^{\alpha-1}  \dot{H}       }{   H^{2} - \Psi_{\alpha} H_{0}^{2} \left (  \frac{H}{H_{0}} \right )^{\alpha}   }    
         =   - \frac{ \alpha \Psi_{\alpha}  \left ( \frac{H}{H_{0}} \right )^{\alpha-2}  \frac{\dot{H}}{H}  }{  1 - \Psi_{\alpha} \left (  \frac{H}{H_{0}} \right )^{\alpha-2}   }                       \notag \\
        &=  - \frac{ \alpha \Psi_{\alpha}   H^{\prime}                                                                     }{  \left (  \frac{H}{H_{0}} \right )^{2-\alpha}    - \Psi_{\alpha}   }  ,
\label{eq:Q_power}
\end{align}
where $\dot{H} = H H^{\prime}$ is also used.
In addition, substituting Eq.\ (\ref{eq:Sol_HH0_power_3}) into Eq.\ (\ref{eq:Q_power}) yields
\begin{align}
    Q  &=  - \frac{ \alpha \Psi_{\alpha}   H^{\prime}                                             }{  (1- \Psi_{\alpha})   e^{- \beta \eta}   + \Psi_{\alpha}      - \Psi_{\alpha}   }  
          = - \frac{ \alpha \Psi_{\alpha} e^{\beta \eta}              H^{\prime}              }{   1- \Psi_{\alpha}                                                                                 }    .
\label{eq:Q_power2}
\end{align}
Dividing Eq.\ (\ref{eq:Q_power2}) by $H$ allows us to write $Q/H$ as  
\begin{align}
    \frac{Q}{H}        & = - \frac{ \alpha \Psi_{\alpha} e^{\beta \eta}                             }{   1- \Psi_{\alpha}                                                                                 }   \frac{  H^{\prime} }{ H }      .                                                                  
\label{eq:QH_power}
\end{align}
Second, we calculate $Q^{\prime}$ to obtain $Q^{\prime}/H$.
Differentiating Eq.\ (\ref{eq:Q_power2}) with respect to $\eta$ yields
\begin{align}  
     Q^{\prime}    &= \frac{d Q}{d  \eta}                                              
                           = \left (  - \frac{ \alpha \Psi_{\alpha} }{  1- \Psi_{\alpha}   }     \right )   ( \beta e^{\beta \eta}   H^{\prime} +  e^{\beta \eta}   H^{\prime \prime}        )    \notag \\
                         &= \left (  - \frac{ \alpha \Psi_{\alpha} }{  1- \Psi_{\alpha}   }     \right )   e^{\beta \eta}  ( \beta   H^{\prime} +   H^{\prime \prime}                              )    .
\label{eq:Qp_power}
\end{align}
Dividing Eq.\ (\ref{eq:Qp_power}) by $H$ gives
\begin{align}  
    \frac{ Q^{\prime} }{H}   &= \left (  - \frac{ \alpha \Psi_{\alpha} }{  1- \Psi_{\alpha}   }     \right )   e^{\beta \eta}  \left ( \beta   \frac{ H^{\prime} }{H} +  \frac{ H^{\prime \prime} }{H}           \right                  )                    .                   
\label{eq:QpH_power}
\end{align}
To calculate this equation, $H^{\prime} / H$ and $H^{\prime \prime} / H$ are required.
The first term $H^{\prime} / H$ is given by Eq. (\ref{eq:Hp-H_power_a_eta_1}), and the second term $H^{\prime \prime} / H$ can be calculated as follows.
After differentiating $H^{\prime}/H_{0}$ [Eq.\ (\ref{eq:Hp-H0_power})] with respect to $\eta$, dividing the resultant equation by $H/H_{0}$ [Eq.\ (\ref{eq:Sol_HH0_power_3})], and performing several calculations, we have
\begin{align}
 \frac{ H^{\prime \prime} }{ H }  &=     \frac{    9 ( 1-  \Psi_{\alpha} )  [ (1-  \Psi_{\alpha})  + (2 -\alpha) \Psi_{\alpha}   e^{\beta \eta}  ]       } {   4 \left [   (1-  \Psi_{\alpha})      +  \Psi_{\alpha} e^{\beta \eta}    \right ]^{2}      }      .
\label{eq:HppH_1}
\end{align}
Thus, $Q^{\prime}/H$ is calculated from Eq.\ (\ref{eq:QpH_power}) by applying Eqs.\ (\ref{eq:Hp-H_power_a_eta_1}) and (\ref{eq:HppH_1}). 
Third, we calculate $f_{\Lambda} (t) / H^{2}$.
Substituting Eq.\ (\ref{eq:fhB(L)_2}) into $f_{\Lambda} (t) / H^{2}$ and applying  Eq.\ (\ref{eq:Sol_HH0_power_3}) gives
\begin{align}
   \frac{  f_{\Lambda} (t) }{  H^{2} }   &= \frac{  \Psi_{\alpha} H_{0}^{2} \left (  \frac{H}{H_{0}} \right )^{\alpha}  }{  H^{2} } 
                                                       =   \Psi_{\alpha}  \left (  \frac{H}{H_{0}} \right )^{\alpha-2}                                                                                        \notag \\
                                                     &=   \frac{ \Psi_{\alpha}                       }{   (1- \Psi_{\alpha})   e^{- \beta \eta}   + \Psi_{\alpha}                           }    
                                                       =   \frac{\Psi_{\alpha} e^{ \beta \eta}  }{   (1- \Psi_{\alpha})                             + \Psi_{\alpha} e^{ \beta \eta}    }       .
\label{eq:3f2H2_00}
\end{align}
In this way, the three terms $Q / H$, $Q^{\prime} / H$, and $f_{\Lambda} (t) / H^{2} $ are obtained.

Substituting Eqs.\ (\ref{eq:Hp-H_power_a_eta_1}) and (\ref{eq:QH_power}) into Eq. (\ref{eq:FL(eta)_1}) yields 
\begin{equation}
F_{\Lambda}  (\eta)=     \frac{  (1-  \Psi_{\alpha})  + (4 + 3\alpha) \Psi_{\alpha}   e^{\beta \eta}  }
                                            { 2 \left [   (1-  \Psi_{\alpha})      +  \Psi_{\alpha} e^{\beta \eta}    \right ]      }        .    
\label{eq:F_L(t)_1}
\end{equation}
Substituting Eqs. (\ref{eq:QH_power}), (\ref{eq:QpH_power}), and (\ref{eq:3f2H2_00}) into Eq. (\ref{eq:GL(eta)_1}) and applying Eqs.\ (\ref{eq:Hp-H_power_a_eta_1}) and (\ref{eq:HppH_1}), we have  
\begin{align}
G_{\Lambda} (\eta) =&   - \frac{3}{2}   +    \frac{  3 [\alpha (2+ \beta) +1]  \Psi_{\alpha}   e^{\beta \eta}                                                         } {   2 \left [   (1-  \Psi_{\alpha})      +  \Psi_{\alpha} e^{\beta \eta}    \right ]           }          \notag \\
                               &  -    \frac{    9 \alpha \Psi_{\alpha} e^{ \beta \eta} [(1-  \Psi_{\alpha})  + (2 -\alpha) \Psi_{\alpha}   e^{\beta \eta}  ]       } {   4 \left [   (1-  \Psi_{\alpha})      +  \Psi_{\alpha} e^{\beta \eta}    \right ]^{2}      }      .
\label{eq:G_G_L(t)_1}
\end{align}
Here, $\beta$ is $\frac{3(2-\alpha)}{2}$ given by Eq.\ (\ref{eq:beta_0}) and $\alpha$ is treated as a real number.
When $\alpha$ was an integer, such as $0$ or $1$, $\Lambda(t)$-$H^{\alpha}$ models were examined although they were considered to be different models.
For example, $\Lambda(t)$-$H^{0}$ models (i.e., $\Lambda$CDM models) and $\Lambda(t)$-$H^{1}$ models were examined as two different models in a previous work \cite{Koma6}.
In the present study, we systematically examine the $\Lambda(t)$-$H^{\alpha}$ model through the free parameter $\alpha$.
(We have confirmed that the above equations are equivalent to results examined in the previous work, when $\alpha =0$ and $\alpha =1$.)

Using $F_{\Lambda} (\eta)$ and $G_{\Lambda} (\eta)$, we numerically solve the differential equation [Eq.\ (\ref{eq:delta-eta_L(t)_1})] for the $\Lambda(t)$-$H^{\alpha}$ model.
To solve this, we use the initial conditions of the Einstein--de Sitter growing model \cite{Lima2011}. 
The initial conditions are given by
\begin{align}
 \delta (\tilde{a}_{i}) = \tilde{a}_{i}  \quad \textrm{and} \quad  \delta^{\prime}  (\tilde{a}_{i}) = \tilde{a}_{i}, 
\label{eq:ICforSolve}
\end{align}
where $\tilde{a}_{i}$ is set to $10^{-3}$ \cite{Lima2011,Koma6}. 
Note that the initial conditions are applied to BV-$H^{\alpha}$ models as well.

\subsubsection{Density perturbations in the BV-$H^{\alpha}$ model} 
\label{DensityBVHA}

Here, we examine density perturbations in the BV-$H^{\alpha}$ model.
The driving term $h_{\textrm{B}}(t)$ given by Eq.\ (\ref{eq:fhB(BV)}) is written as 
\begin{equation}
       h_{\textrm{B}}(t)  =   \frac{3}{2}  \Psi_{\alpha} H_{0}^{2} \left (  \frac{H}{H_{0}} \right )^{\alpha}    .
\label{eq:fhB(BV)2}
\end{equation}
Then, we rewrite Eqs.\ (\ref{eq:delta-eta_c=0_BV_1})--(\ref{eq:GB(eta)_c=0_1}).
\begin{equation}
\delta^{\prime \prime}  + F_{\textrm{B}} (\eta) \delta^{\prime}  +  G_{\textrm{B}} (\eta) \delta =0, 
\label{eq:delta-eta_c=0_BV_2}
\end{equation}
\begin{equation}
F_{\textrm{B}} (\eta) =  2  + \frac{ \Gamma + H^{\prime} }{ H }   -   \frac{  \Gamma H^{\prime} - H \Gamma^{\prime}  }{ H (3H -\Gamma) }   ,
\label{eq:FB(eta)_c=0_2}
\end{equation}
\begin{equation}
G_{\textrm{B}} (\eta)  =  \left (  \frac{ \Gamma }{ H }  - 1   \right )   \left (  \frac{ \Gamma }{ 2H }  + \frac{3}{2}   \right )   -   \frac{ 3 (\Gamma H^{\prime} - H \Gamma^{\prime})  }{ H (3H -\Gamma) }   ,
\label{eq:GB(eta)_c=0_2}
\end{equation}
where, from Eq.\ (\ref{eq:gamma_BV}), $\Gamma /H$ can be written as 
\begin{equation}
     \frac{  \Gamma }{ H }  =    \frac{3}{4 \pi G}    \frac{ h_{\textrm{B}}(t) }{ \rho }                , 
\label{eq:Gamma-H_w=0_hb}
\end{equation}
and $\rho$ is given by
\begin{equation}
 \rho = \frac{3 H^{2} }{ 8\pi G }              .
\label{eq:rho_BV}
\end{equation}
Equation\ (\ref{eq:rho_BV}) is obtained from the Friedmann equation by substituting $f_{\Lambda}(t) =  0$ into Eq.\ (\ref{eq:General_FRW01}).

We now calculate $F_{\textrm{B}} (\eta)$ and $G_{\textrm{B}} (\eta)$.
To this end, we require three terms, namely $\Gamma / H$, $\Gamma^{\prime} / H$, and $\frac{  \Gamma H^{\prime} - H \Gamma^{\prime}  }{ H (3H -\Gamma) }$.
The second term is required for calculating the third term.
First, we calculate $\Gamma / H$.
Substituting Eqs.\ (\ref{eq:fhB(BV)2}) and (\ref{eq:rho_BV}) into Eq.\ (\ref{eq:Gamma-H_w=0_hb}) yields  
\begin{align}
     \frac{  \Gamma }{ H }  =    \frac{3}{4 \pi G}  \frac{  \frac{3}{2}  \Psi_{\alpha}  H_{0}^{2} \left (  \frac{H}{H_{0}} \right )^{\alpha}   }{ \frac{3 H^{2} }{ 8\pi G }  }    
                                      &=  3 \Psi_{\alpha} \left (  \frac{H}{H_{0}} \right )^{\alpha -2}         ,
\label{eq:Gamma-H_power}
\end{align}
or equivalently,
\begin{align}
       \Gamma      &=  3 \Psi_{\alpha} H_{0} \left (  \frac{H}{H_{0}} \right )^{\alpha -1}         .
\label{eq:Gamma_power}
\end{align}
Substituting Eq.\ (\ref{eq:Sol_HH0_power_3}) into Eq.\ (\ref{eq:Gamma-H_power}) yields 
\begin{align}
     \frac{  \Gamma }{ H }  &=  3 \Psi_{\alpha} \left (  \frac{H}{H_{0}} \right )^{\alpha -2}       
                                        =  \frac{ 3 \Psi_{\alpha}                                } { (1- \Psi_{\alpha})  e^{- \beta \eta}  + \Psi_{\alpha}                         }    \notag \\                          
                                      &=  \frac{ 3 \Psi_{\alpha}  e^{\beta \eta}         } { (1- \Psi_{\alpha})                           + \Psi_{\alpha} e^{\beta \eta}   }    .
\label{eq:Gamma-H_power_a_eta}
\end{align}
Second, we calculate $\Gamma^{\prime} / H$, which is used for determining the third term.
After differentiating Eq.\ (\ref{eq:Gamma_power}) with respect to $\eta$, applying Eq.\ (\ref{eq:Gamma-H_power_a_eta}), and dividing the resultant equation by $H$, we write $\Gamma^{\prime} / H$ as
\begin{equation}
   \frac{ \Gamma^{\prime} }{H} = (\alpha -1) \frac{\Gamma}{H}  \frac{ H^{\prime} }{H}   .
\label{eq:Gammaprime-H_power}
\end{equation}
Third, we calculate $\frac{  \Gamma H^{\prime} - H \Gamma^{\prime}  }{ H (3H -\Gamma) }$.
Reformulating this term and substituting Eq.\ (\ref{eq:Gammaprime-H_power}) into the resultant equation yields
\begin{align}
 \frac{  \Gamma H^{\prime} - H \Gamma^{\prime}  }{ H (3H -\Gamma) }  &=  \frac{  \frac{\Gamma}{H} \frac{ H^{\prime} }{H} -    \frac{ \Gamma^{\prime} }{H} }{ 3 -  \frac{\Gamma}{H} } 
                                                                                                          =  \frac{  \frac{\Gamma}{H} \frac{ H^{\prime} }{H} -    (\alpha -1) \frac{\Gamma}{H}  \frac{ H^{\prime} }{H}  }{ 3 -  \frac{\Gamma}{H} }  \notag  \\
                                                                                                        &=  \frac{  (2- \alpha) \frac{\Gamma}{H} \frac{ H^{\prime} }{H} }{ 3 -  \frac{\Gamma}{H} }  .
\label{eq:Complex_power_0}
\end{align}
Substituting Eqs.\ (\ref{eq:Hp-H_power_a_eta_1}) and (\ref{eq:Gamma-H_power_a_eta}) into Eq.\ (\ref{eq:Complex_power_0}) gives
\begin{align}
 \frac{  \Gamma H^{\prime} - H \Gamma^{\prime}  }{ H (3H -\Gamma) }      &=   \frac{ -\frac{3}{2} (2-\alpha) \Psi_{\alpha}   e^{\beta \eta}  }{   (1-  \Psi_{\alpha})      +  \Psi_{\alpha} e^{\beta \eta}    }  .
\label{eq:Complex_power_1}
\end{align}
From these results, we can calculate $F_{\textrm{B}} (\eta)$ and $G_{\textrm{B}} (\eta)$.
Substituting Eqs. (\ref{eq:Hp-H_power_a_eta_1}),  (\ref{eq:Gamma-H_power_a_eta}), and (\ref{eq:Complex_power_1}) into Eq. (\ref{eq:FB(eta)_c=0_2}) yields 
\begin{equation}
F_{\textrm{B}} (\eta)=     \frac{  (1-  \Psi_{\alpha})  + (16 - 3\alpha) \Psi_{\alpha}   e^{\beta \eta}  }
                                            { 2 \left [   (1-  \Psi_{\alpha})      +  \Psi_{\alpha} e^{\beta \eta}    \right ]      }        .    
\label{eq:F_BV-power_c0_1}
\end{equation}
Substituting Eqs. (\ref{eq:Gamma-H_power_a_eta}) and (\ref{eq:Complex_power_1}) into Eq. (\ref{eq:GB(eta)_c=0_2}) yields 
\begin{align}
G_{\textrm{B}} (\eta) =&   - \frac{3}{2}   +    \frac{   (24 - 9\alpha) \Psi_{\alpha}   e^{\beta \eta}  } {   2 \left [   (1-  \Psi_{\alpha})      +  \Psi_{\alpha} e^{\beta \eta}    \right ]           }          \notag \\
                                 &                       +    \frac{    9 \Psi_{\alpha}^{2}   e^{2 \beta \eta}           } {   2 \left [   (1-  \Psi_{\alpha})      +  \Psi_{\alpha} e^{\beta \eta}    \right ]^{2}      }      .
\label{eq:G_BV-power_c0_1}
\end{align}
Here, $\beta$ is $\frac{3(2-\alpha)}{2}$ given by Eq.\ (\ref{eq:beta_0}), and $\alpha$ is treated as a real number.
Using $F_{\textrm{B}} (\eta)$, $G_{\textrm{B}} (\eta)$, and the initial conditions given by Eq.\ (\ref{eq:ICforSolve}), we can numerically solve the differential equation [Eq.\ (\ref{eq:delta-eta_c=0_BV_2})] for the BV-$H^{\alpha}$ model.

When $\alpha$ was an integer, such as $0$ or $1$, the BV-$H^{\alpha}$ models were examined, although they were considered to be different models.
For example, BV-$H^{0}$ and BV-$H^{1}$ models were examined as two different models \cite{Koma6}.
We have confirmed that Eqs.\ (\ref{eq:F_BV-power_c0_1}) and (\ref{eq:G_BV-power_c0_1}) are equivalent to those in Ref.\ \cite{Koma6} when $\alpha =0$ and $\alpha =1$.
Cosmological models similar to the BV-$H^{0}$ and BV-$H^{1}$ models were investigated in, for example, Refs.\ \cite{Lima2011} and \cite{Barrow21}, respectively.
In this study, $\alpha$ is a free parameter and a real number. Therefore, we can systematically examine the BV-$H^{\alpha}$ model, which was not possible in previous works.

\subsection{Thermodynamic constraints}
\label{Thermodynamic constraints}

Ordinary, isolated macroscopic systems spontaneously evolve to equilibrium states that maximize the entropy consistent with their constraints \cite{Callen}.
In other words, the entropy of such systems does not decrease (i.e., the second law of thermodynamics) and approaches a certain maximum value at the last stage (i.e., the maximization of entropy) \cite{Koma14}.
In fact, a certain type of universe is expected to be constrained by thermodynamics as if it behaves as the macroscopic system \cite{Mimoso2013}.

In this subsection, we use the results of previous works \cite{Koma14,Koma15} to review such thermodynamic constraints on the $\Lambda (t)$-$H^{\alpha}$ and BV-$H^{\alpha}$ models.
In fact, the two models always satisfy the second law of thermodynamics, whereas they satisfy the maximization of entropy only under specific conditions \cite{Koma14,Koma15}.
In particular, the maximization of entropy depends almost entirely on the constraints on $\ddot{S}_{\rm{BH}} <0$, where $S_{\rm{BH}}$ is the Bekenstein--Hawking entropy \cite{Koma15}.
Therefore, to discuss the thermodynamic constraints, we examine $\ddot{S}_{\rm{BH}} <0$.
For this purpose, we present the Bekenstein--Hawking entropy on the horizon of the universe.

The Bekenstein--Hawking entropy $S_{\rm{BH}}$ is written as 
\begin{equation}
 S_{\rm{BH}}  = \frac{ k_{B} c^3 }{  \hbar G }  \frac{A_{H}}{4}   ,
\label{eq:SBH}
\end{equation}
where $\hbar$ is the reduced Planck constant, defined as $\hbar \equiv h/(2 \pi)$ using the Planck constant $h$ \cite{Koma11,Koma12,Koma14} and 
$A_{H}$ is the surface area of a sphere with a Hubble horizon $r_{H}$, which is given by $c/H$.
In a flat FRW universe, the Hubble horizon is equivalent to the apparent horizon.
Substituting $A_{H}=4 \pi r_{H}^2 $ into Eq.\ (\ref{eq:SBH}) and applying $r_{H} =c/H$ yields 
\begin{equation}
S_{\rm{BH}} =  \left ( \frac{ \pi k_{B} c^5 }{ \hbar G } \right )  \frac{1}{H^2}     .
\label{eq:SBH_1}      
\end{equation}
This equation indicates that $S_{\rm{BH}}$ depends on the background evolution of the universe.
Therefore, the evolution of $S_{\rm{BH}}$ in the two models is equivalent when their background evolution is the same.

Substituting Eq.\ (\ref{eq:Sol_HH0_power}) into Eq.\ (\ref{eq:SBH_1}) and performing several calculation, we can obtain $S_{\rm{BH}}$, $\dot{S}_{\rm{BH}}$, and $\ddot{S}_{\rm{BH}}$.
The result of Ref.\ \cite{Koma15} is used to write the normalized $\ddot{S}_{\rm{BH}}$ as 
\begin{equation}  
   \frac{ \ddot{S}_{\rm{BH}}  }{S_{\rm{BH},0} H_{0}^{2} } =      
                                                               \frac{9}{2}    \frac{  (1- \Psi_{\alpha})    \tilde{a}^{ - \beta }     \left [  (1- \Psi_{\alpha})    \tilde{a}^{ - \beta }    + (\alpha-2)\Psi_{\alpha}   \right ]             }{   \left [    (1- \Psi_{\alpha})   \tilde{a}^{ - \beta  }  + \Psi_{\alpha}         \right ]^{2}      }        ,
\label{eq:d2SBH2SBH0_power_1}
\end{equation}
where $S_{\rm{BH},0}$ is the current Bekenstein--Hawking entropy and $\beta$ is $ \frac{3(2-\alpha)}{2}$ given by Eq.\ (\ref{eq:beta_0}). 
Equation\ (\ref{eq:d2SBH2SBH0_power_1}) indicates that $ \ddot{S}_{\rm{BH}} < 0$ should be satisfied at least in the last stage, that is, $ \tilde{a} \rightarrow \infty$, when $\alpha < 2$ \cite{Koma14}.
In other words, a region that satisfies $ \ddot{S}_{\rm{BH}} < 0$ in the $(\Psi_{\alpha}, \alpha)$ plane varies with time before the last stage.
To study such a relaxation-like process, we use the boundary required for $\ddot{S}_{\rm{BH}} = 0$, which is given by \cite{Koma14}
\begin{align}  
   \Psi_{\alpha} =  \frac{  \tilde{a}^{ - \beta  }       }{  2 - \alpha +  \tilde{a}^{ - \beta  }       }  .
\label{eq:d2Sdt2_power_0_aa0_3}
\end{align}
From this equation, we can plot the boundary of $\ddot{S}_{\rm{BH}} = 0$ in the $(\Psi_{\alpha}, \alpha)$ plane.
Here $\Psi_{\alpha}$ corresponds to a density parameter for effective dark energy.

\begin{figure} [t] 
\begin{minipage}{0.495\textwidth}
\begin{center}
\scalebox{0.3}{\includegraphics{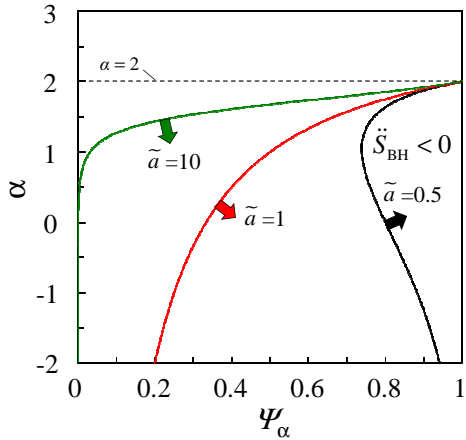}}
\end{center}
\end{minipage}
\caption{ (Color online).  Thermodynamic constraints on the $\Lambda (t)$-$H^{\alpha}$ and BV-$H^{\alpha}$ models in the $(\Psi_{\alpha}, \alpha)$ plane.
The boundary of $\ddot{S}_{\rm{BH}} = 0$ for $\tilde{a}=0.5$, $1$, and $10$ is shown.
The boundary for the two models is the same because the same background evolution is used for both models.
The arrow on each boundary indicates a region that satisfies $\ddot{S}_{\rm{BH}} < 0$.
All the boundaries intersect at the point $(\Psi_{\alpha}, \alpha) = (1, 2)$.
The horizontal dashed line represents $\alpha=2$.
The region below the dashed line should satisfy $\ddot{S}_{\rm{BH}} < 0$, at least in the last stage \cite{Koma14}.
In the last stage, the region should also satisfy observational constraints on an initially decelerating and then accelerating universe \cite{Koma15}.
Similar thermodynamic constraints have been examined in Refs.\ \cite{Koma14,Koma15}. }
\label{Fig-dS2dt2_plane_power}
\end{figure}

To examine the evolution of the boundary, typical boundaries for $\tilde{a}=0.5$, $1$, and $10$ are shown in Fig.\ \ref{Fig-dS2dt2_plane_power}.
The boundary for the $\Lambda (t)$-$H^{\alpha}$ and BV-$H^{\alpha}$ models is the same because the background evolution of the universe in both models is equivalent.
The arrow on each boundary indicates a region that satisfies $\ddot{S}_{\rm{BH}} < 0$.
(Similar boundaries are examined in Refs.\ \cite{Koma14,Koma15}.)
As shown in Fig.\ \ref{Fig-dS2dt2_plane_power}, this region gradually extends and approaches $\alpha =2$ with increasing $\tilde{a}$. 
When $\alpha < 2$, maximization of the entropy, $\ddot{S}_{\rm{BH}} < 0$, should be satisfied, at least in the last stage of the evolution of an expanding universe \cite{Koma14}.
In this way, thermodynamic constraints on the two models can be discussed in the $(\Psi_{\alpha}, \alpha)$ plane.
In the next section, we examine observational constraints in combination with the thermodynamic constraints shown here.

It should be noted that cosmological adiabatic particle creation results in the generation of irreversible entropy \cite{Prigogine_1988-1989}. 
The irreversible entropy $S_{m}$ due to adiabatic particle creation was examined in a previous work \cite{Koma15}.
Consequently, $ \ddot{S}_{m} < 0 $ is found to be always satisfied when $\alpha < 2$.
That is, constraints on $\ddot{S}_{m}<0$ are slightly looser than those on $\ddot{S}_{\rm{BH}}<0$.
In addition, it is well known that $S_{\rm{BH}}$ is extremely large in comparison with the other entropies \cite{Egan1}.
These results indicate that the maximization of entropy depends almost entirely on the constraints on $\ddot{S}_{\rm{BH}} <0$, as examined in Ref.\ \cite{Koma15}.
Accordingly, we use $\ddot{S}_{\rm{BH}} <0$ to discuss the thermodynamic constraints in this study.

\section{Evolution of the universe in the $\Lambda(t)$-$H^{\alpha}$ and BV-$H^{\alpha}$ models}
\label{Evolution of the universe}

In this section, we examine the evolution of the universe in the $\Lambda(t)$-$H^{\alpha}$ and BV-$H^{\alpha}$ models.
In Sec.\ \ref{Evolution of the universe for Psi=0.685}, the evolution of the universe for $\Psi_{\alpha}=0.685$ is discussed as a specific case.
In Sec.\ \ref{Observational and thermodynamic constraints}, the observational and thermodynamic constraints on the two models are investigated with chi-squared functions in the $(\Psi_{\alpha}, \alpha)$ plane.
Here, $\Psi_{\alpha}$ and $\alpha$ are treated as free parameters.
Note that we do not discuss the significant tension between the Planck results \cite{Planck2018} and the local (distance ladder) measurement from the Hubble Space Telescope \cite{Riess20162019}.

\subsection{Evolution of the universe for $\Psi_{\alpha} = 0.685$}
\label{Evolution of the universe for Psi=0.685}

We examine the typical evolution of the universe in the $\Lambda(t)$-$H^{\alpha}$ and BV-$H^{\alpha}$ models.
To this end, $\Psi_{\alpha}$ is set to $0.685$, which is equivalent to $\Omega_{\Lambda}$ for the standard $\Lambda$CDM model from the Planck 2018 results \cite{Planck2018}.
($\Psi_{\alpha}$ corresponds to a density parameter for effective dark energy.)
Therefore, the following result for $\alpha =0$ of the $\Lambda (t)$-$H^{\alpha}$ model is equivalent to that for the $\Lambda$CDM model.

\subsubsection{Background evolution of the universe for $\Psi_{\alpha} = 0.685$}
\label{Background evolution}

\begin{figure} [t] 
\begin{minipage}{0.495\textwidth}
\begin{center}
\scalebox{0.31}{\includegraphics{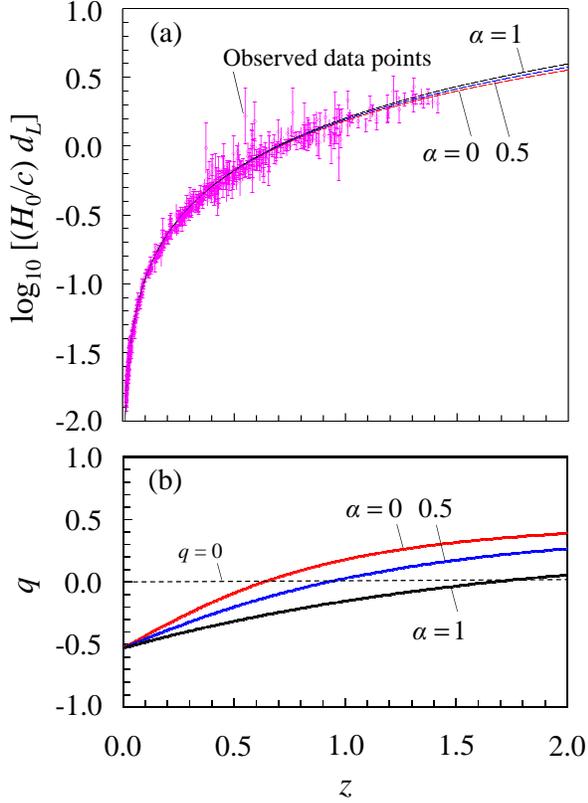}}
\end{center}
\end{minipage}
\caption{ (Color online). Background evolution of the universe in the $\Lambda (t)$-$H^{\alpha}$ and BV-$H^{\alpha}$ models for $\Psi_{\alpha} =0.685$.
(a) Luminosity distance $d_L$.
(b) Deceleration parameter $q$.
The background evolution of the universe in the two models is the same.
In (a), the symbols with error bars are observed supernova data points taken from Ref.\ \cite{Suzuki_2012}.
To normalize the data points, $H_{0}$ is set to $67.4$ km/s/Mpc from the Planck 2018 results \cite{Planck2018}.  
In (b), the horizontal dashed line represents $q=0$.
  }
\label{Fig-dL-z_power}
\end{figure}

To examine the background evolution of the universe, we use the luminosity distance $d_{L}$ \cite{Sato1,Carroll01}, which is written as  
\begin{equation}
  \left ( \frac{ H_{0} }{ c } \right )   d_{L}      =   (1+z)  \int_{1}^{1+z}  \frac{dy} { F(y) }    ,
\label{eq:dL}  
\end{equation}
where the redshift $z$ is given by
\begin{equation}
       z = \tilde{a}^{-1} -1 .
\label{eq:z}
\end{equation}
The integrating variable $y$ and the function $F(y)$ are given by 
\begin{equation}
   y =  \tilde{a}^{-1}  \quad \textrm{and} \quad    F(y)  = \frac{ H }{ H_{0} } .
\end{equation}
For the $\Lambda(t)$-$H^{\alpha}$ and BV-$H^{\alpha}$ models, $H/H_{0}$ is given by Eq.\ (\ref{eq:Sol_HH0_power}).
The background evolution of the universe in the $\Lambda (t)$-$H^{\alpha}$ and BV-$H^{\alpha}$ models is equivalent and therefore, $d_L$ for the two models is the same.
Similarly, the temporal deceleration parameter $q$ for the two models is the same, where $q$ is given by Eq.\ (\ref{eq:q_power}).

Figure\ \ref{Fig-dL-z_power} shows the background evolution of the universe for $\Psi_{\alpha} =0.685$.
To examine typical results, $\alpha$ is set to $0$, $0.5$, and $1$.
In Fig.\ \ref{Fig-dL-z_power}(a), the observed data points are the \textit{Union 2.1} set of $580$ type Ia supernovae \cite{Suzuki_2012}.
As shown in Fig.\ \ref{Fig-dL-z_power}(a), the luminosity distance $d_{L}$ for $\alpha =0$, $0.5$, and $1$ is likely consistent with the supernova data.
The luminosity distance $d_{L}$ for $\alpha = 0.5$ and $1$ deviates from $d_{L}$ for $\alpha = 0$.
Note that the deviation looks small because a logarithmic scale is used for the vertical axis in this figure.
From Fig.\ \ref{Fig-dL-z_power}(b), we can confirm that $\alpha$ affects the evolution of the deceleration parameter $q$.
Of course, both $\Psi_{\alpha}$ and $\alpha$ affect the background evolution of the universe.
We examine this influence later using chi-squared functions in the $(\Psi_{\alpha}, \alpha)$ plane.

\subsubsection{Evolution of density perturbations for $\Psi_{\alpha} = 0.685$}
\label{Evolution of density perturbations}

In this study, the background evolution of the universe is the same in both models.
However, even with this similarity, the density perturbations are expected to be different.
In this subsection, we examine first-order density perturbations in the $\Lambda(t)$-$H^{\alpha}$ and BV-$H^{\alpha}$ models for $\Psi_{\alpha} =0.685$.  

We first examine the evolution of the perturbation growth factor $\delta$ for the two models.
In this study, $\delta$ is numerically solved using the initial conditions given by Eq.\ (\ref{eq:ICforSolve}).
To examine typical results, $\alpha$ is set to $0$, $0.5$, and $1$.
Accordingly, the background evolution considered here is the same as that shown in Fig.\ \ref{Fig-dL-z_power}.
A previous work \cite{Koma6} investigated similar density perturbations using the $\Lambda(t)$-$H^{0}$, $\Lambda(t)$-$H^{1}$, BV-$H^{0}$, and BV-$H^{1}$ models.
In this study, $\alpha$ is a free parameter that can be treated as a real number.
Therefore, we can systematically examine the difference between the $\Lambda(t)$-$H^{\alpha}$ and BV-$H^{\alpha}$ models.

\begin{figure} [b] 
\begin{minipage}{0.495\textwidth}
\begin{center}
\scalebox{0.3}{\includegraphics{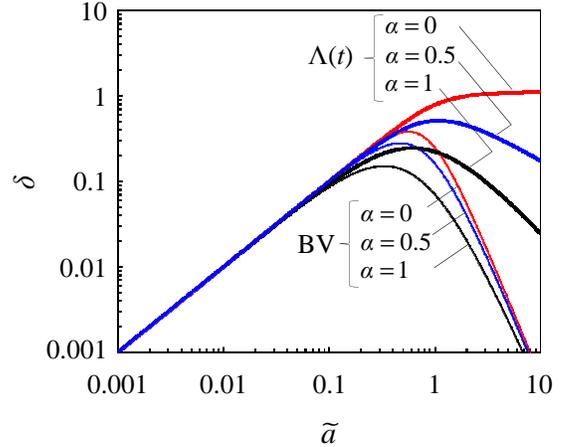}}
\end{center}
\end{minipage}
\caption{ (Color online). Evolution of $\delta$ for the $\Lambda(t)$-$H^{\alpha}$ and BV-$H^{\alpha}$ models with $\Psi_{\alpha} =0.685$.
The bold and thin curves represent the $\Lambda(t)$-$H^{\alpha}$ and BV-$H^{\alpha}$ models, respectively.
For the two models, $\alpha$ is set to $0$, $0.5$, and $1$.
 }
\label{Fig-delta-a_power}
\end{figure}

For $\tilde{a} \lessapprox 0.1$, $\delta$ increases with $\tilde{a}$, as shown in Fig.\ \ref{Fig-delta-a_power}. 
Thereafter, the increase of $\delta$ tends to gradually slow. 
For $\tilde{a} \gtrapprox 1$, $\delta$ for $\alpha =0$ of the $\Lambda (t)$-$H^{\alpha}$ model does not decrease, whereas $\delta$ for the others decreases.
In this way, $\alpha$ and the type of model affect the density perturbations.
In particular, the decrease in $\delta$ for the BV-$H^{\alpha}$ model is significant, in comparison with the $\Lambda(t)$-$H^{\alpha}$ model.
As expected, density perturbations in the two models are greatly different even if the background evolution is the same.

\begin{figure} [t] 
\begin{minipage}{0.495\textwidth}
\begin{center}
\scalebox{0.31}{\includegraphics{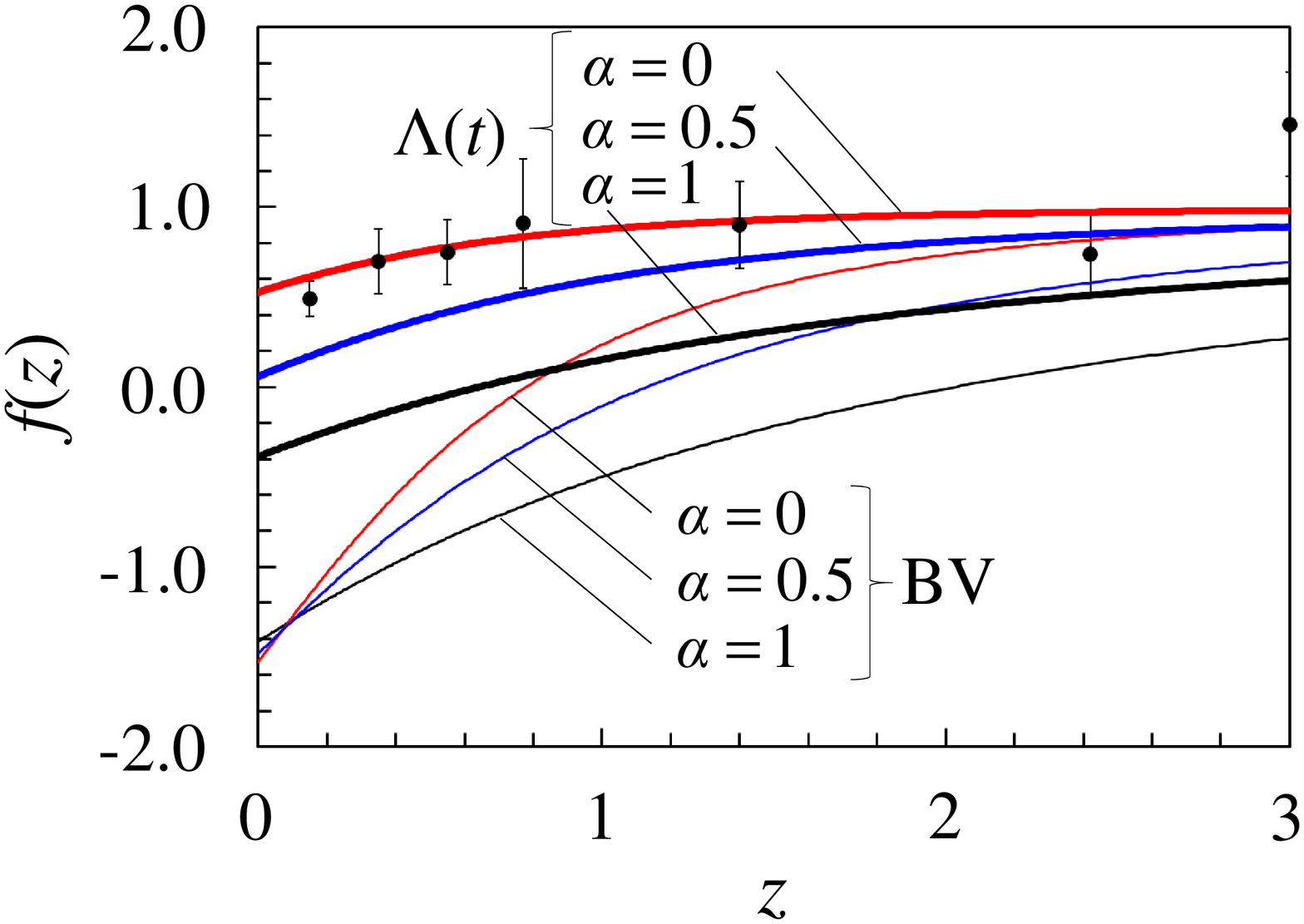}}
\end{center}
\end{minipage}
\caption{ (Color online). Evolution of $f(z)$ for the $\Lambda(t)$-$H^{\alpha}$ and BV-$H^{\alpha}$ models with $\Psi_{\alpha} =0.685$.
The bold and thin curves represent the $\Lambda(t)$-$H^{\alpha}$ and BV-$H^{\alpha}$ models, respectively.
The closed circles with error bars are observed data points taken from a summary in Jesus \textit{et al.} \cite{Lima2011}.
Each original data point is given in Refs.\ \cite{Colless2001,Guzzo2008,Tegmark2006,Ross2007,Angela2007,Viel2004,McDonald2005}. }
\label{Fig-f(z)-z_power}
\end{figure}

\begin{figure} [t] 
\begin{minipage}{0.495\textwidth}
\begin{center}
\scalebox{0.31}{\includegraphics{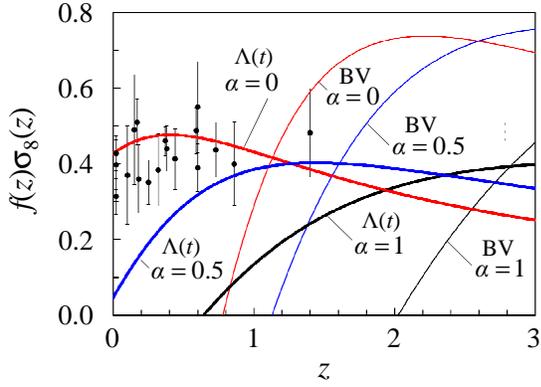}}
\end{center}
\end{minipage}
\caption{ (Color online). Evolution of $f(z) \sigma_{8}(z)$ for the $\Lambda(t)$-$H^{\alpha}$ and BV-$H^{\alpha}$ models with $\Psi_{\alpha} =0.685$.
The bold and thin curves represent the $\Lambda(t)$-$H^{\alpha}$ and BV-$H^{\alpha}$ models, respectively.
The closed circles with error bars are observed data points taken from a summary in Nesseris \textit{et al.} \cite{Nesseris2017}.
Each original data point is given in Refs.\ \cite{58,59,60,61,62,63,67,68,69,71,72,73,74,75}. }
\label{Fig-f(z)sigma8-z_power}
\end{figure}

Next, we use the obtained $\delta$ to calculate an indicator of clustering, namely the growth rate $f(z)$ of clustering \cite{Peebles_1993} given by 
\begin{equation}
 f(z) = \frac{d \ln \delta }{  d \ln a } = - (1 + z ) \frac{d \ln \delta }{  dz }     . 
\label{eq:f(z)}
\end{equation}
In addition, we calculate a combination value $f(z) \sigma_{8}(z)$.
Here $\sigma_{8} (z)$ is the redshift-dependent root-mean-square (rms) fluctuations of the linear density field within a sphere of radius $R = 8h^{-1}$ Mpc \cite{Nesseris2017}
(where $h$ is the reduced Hubble constant defined by $h = H_{0} / 100$).
The redshift-dependent rms fluctuations $\sigma_{8}(z)$ can be written as \cite{Basilakos2014}
\begin{equation}
\sigma_{8}(z) =  \sigma_{8}  \left [ \frac{\delta (z)}{\delta (z=0)} \right ]      , 
\label{eq:sigma8(z)}
\end{equation}
where $\sigma_{8}$ is $\sigma_{8} (z)$ at redshift $z=0$.
We set $\sigma_{8} =0.811$ from the Planck 2018 results \cite{Planck2018}.
Note that more exact formalisms are examined elsewhere, such as Refs.\ \cite{Sola_2015L14,Valent2015,Sola2019}.

The evolution of $f(z)$ and $f(z) \sigma_{8}(z)$ is shown in Figs.\ \ref{Fig-f(z)-z_power} and \ref{Fig-f(z)sigma8-z_power}, respectively.
As shown in Fig.\ \ref{Fig-f(z)-z_power}, $f(z)$ for $\alpha =0$ of the $\Lambda (t)$-$H^{\alpha}$ model agrees with the observed data points, whereas $f(z)$ for the other cases does not.
Similarly, $f(z) \sigma_{8}(z)$ for $\alpha =0$ of the $\Lambda (t)$-$H^{\alpha}$ model agrees with the observed data points (Fig.\ \ref{Fig-f(z)sigma8-z_power}).
In contrast, at low $z$, $f(z)$ and $f(z) \sigma_{8}(z)$ for the BV-$H^{\alpha}$ model disagree with the data points, in comparison with the $\Lambda(t)$-$H^{\alpha}$ model. 
This is because, as shown in Fig.\ \ref{Fig-delta-a_power}, $\delta$ for the BV-$H^{\alpha}$ model decays at large $\tilde{a}$, corresponding to low $z$. 

A previous work \cite{Koma6} discussed similar results using four different models, corresponding to the $\Lambda(t)$-$H^{0}$, $\Lambda(t)$-$H^{1}$, BV-$H^{0}$, and BV-$H^{1}$ models.
In the present study, we can systematically examine the $\Lambda(t)$-$H^{\alpha}$ and BV-$H^{\alpha}$ models using the free parameter $\alpha$.
We discuss the systematic study in the next subsection.

\subsection{Constraints on the $\Lambda (t)$-$H^{\alpha}$ and BV-$H^{\alpha}$ models}
\label{Observational and thermodynamic constraints}

\begin{figure*} [htb] 
 \begin{minipage}{0.495\hsize}
  \begin{center}
   \scalebox{0.3}{\includegraphics{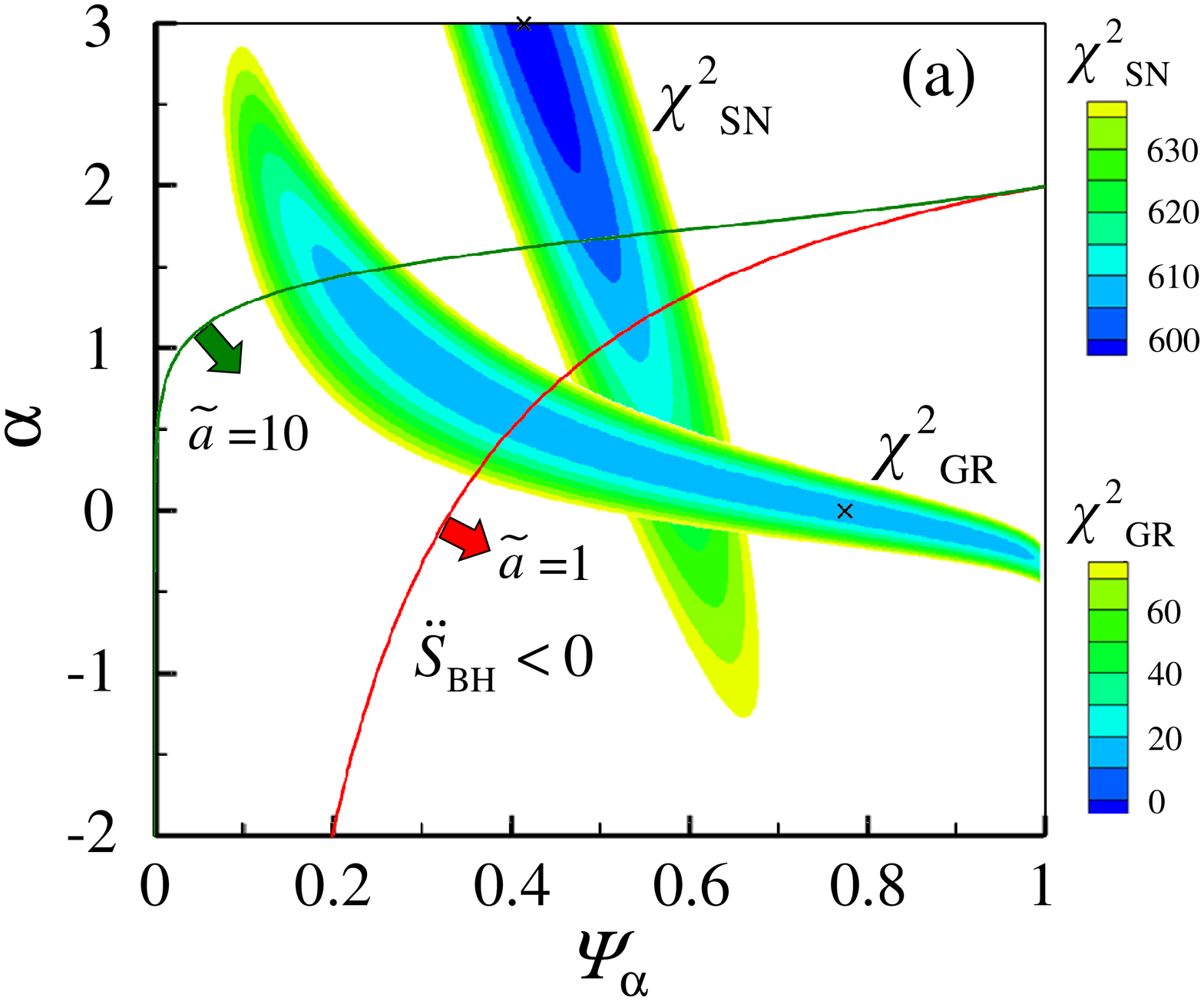}}\\  
 \end{center}
 \end{minipage}
 \begin{minipage}{0.495\hsize}
  \begin{center}
   \scalebox{0.3}{\includegraphics{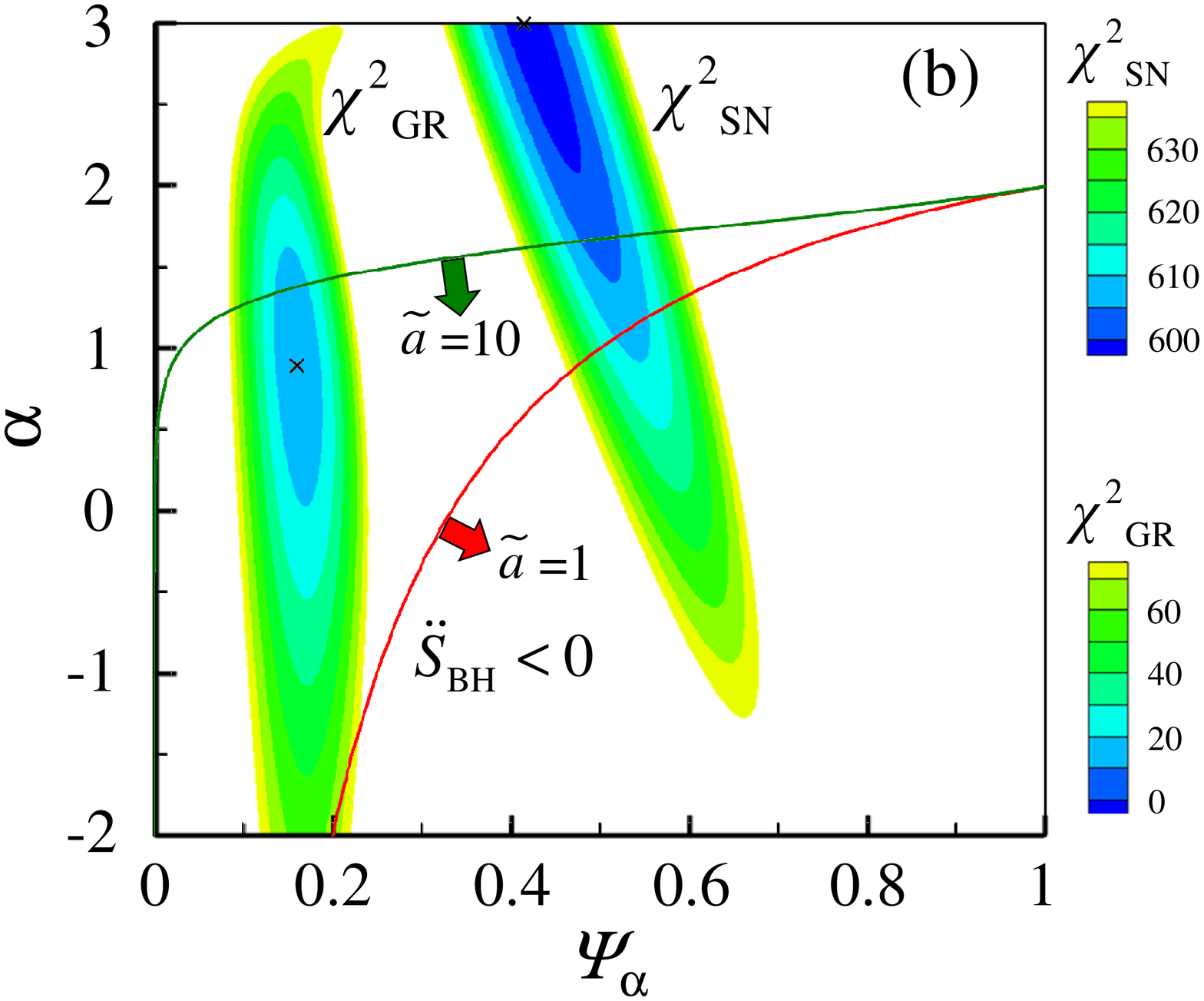}}\\
  \end{center}
 \end{minipage}
\caption{ (Color online). 
Contours of $\chi_{\textrm{SN}}^{2}$ for supernovae and $\chi_{\textrm{GR}}^{2}$ for the growth rate in the $(\Psi_{\alpha}, \alpha)$ plane.
(a) $\Lambda (t)$-$H^{\alpha}$ model.  (b) BV-$H^{\alpha}$ model.
The color scale bars for $\chi_{\textrm{SN}}^{2}$ and $\chi_{\textrm{GR}}^{2}$ are given at the upper and lower sides, respectively.
Small-$\chi_{\textrm{SN}}^{2}$ regions $(\chi_{\textrm{SN}}^{2} < 640)$ and small-$\chi_{\textrm{GR}}^{2}$ regions $(\chi_{\textrm{GR}}^{2} < 80)$ are displayed.
The boundary of $\ddot{S}_{\rm{BH}} = 0$ for $\tilde{a}=1$ and $10$ is plotted from Fig.\ \ref{Fig-dS2dt2_plane_power}.
The arrow on each boundary indicates a region that satisfies the maximization of the entropy, that is, $\ddot{S}_{\rm{BH}} < 0$.
The contours of $\chi_{\textrm{SN}}^{2}$ and the boundary of $\ddot{S}_{\rm{BH}} = 0$ in (a) are the same as those in (b). 
The x's mark the location of the minimum value of each chi-squared function.
The minimum values are summarized in Table\ \ref{tab-chi2min}.  
}
\label{Fig-SN_GFfsigma8_dS2BH=0_plane}
\end{figure*}
\begin{figure*} [htb] 
 \begin{minipage}{0.495\hsize}
  \begin{center}
   \scalebox{0.3}{\includegraphics{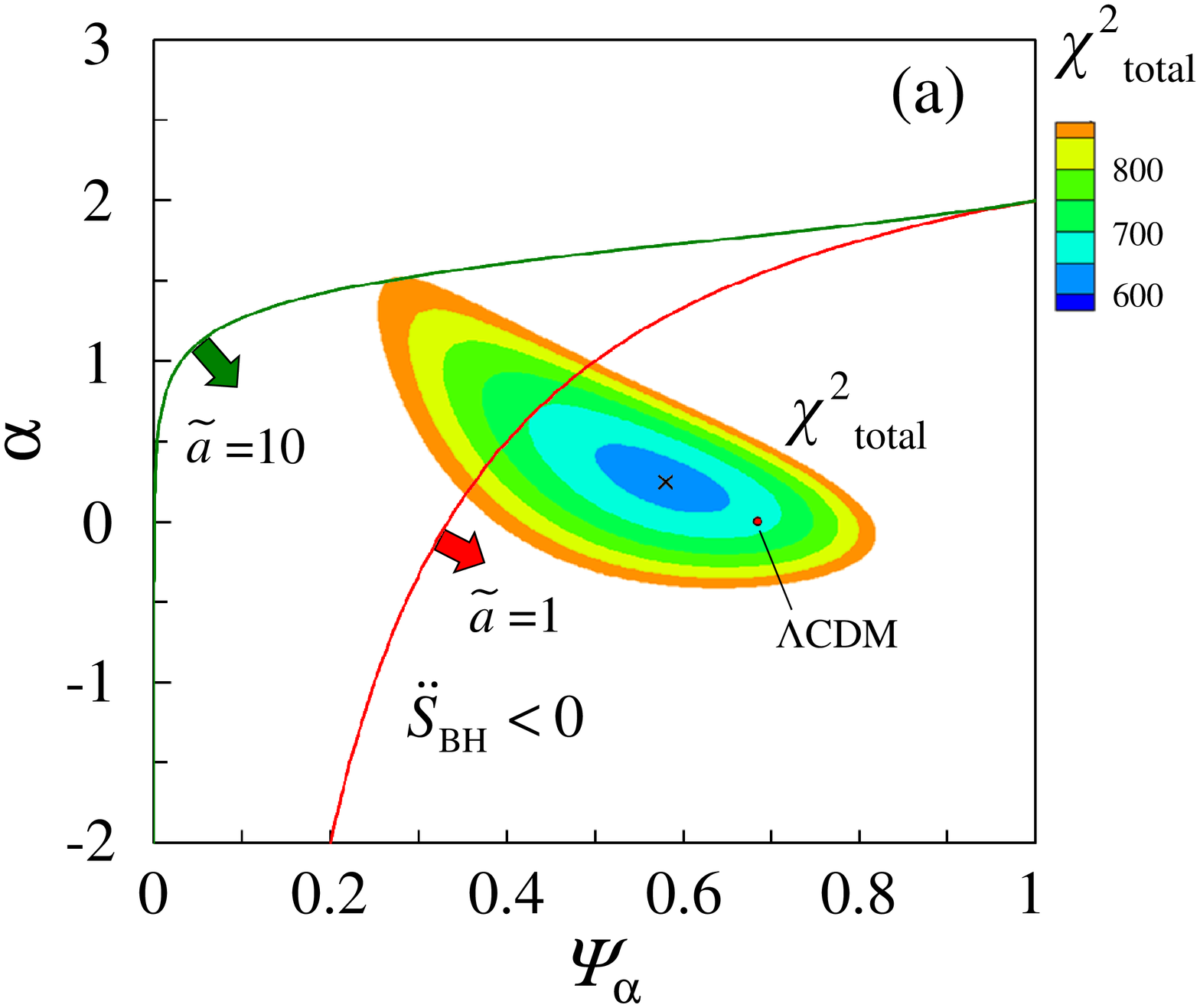}}\\  
  \end{center}
 \end{minipage}
 \begin{minipage}{0.495\hsize}
  \begin{center}
   \scalebox{0.3}{\includegraphics{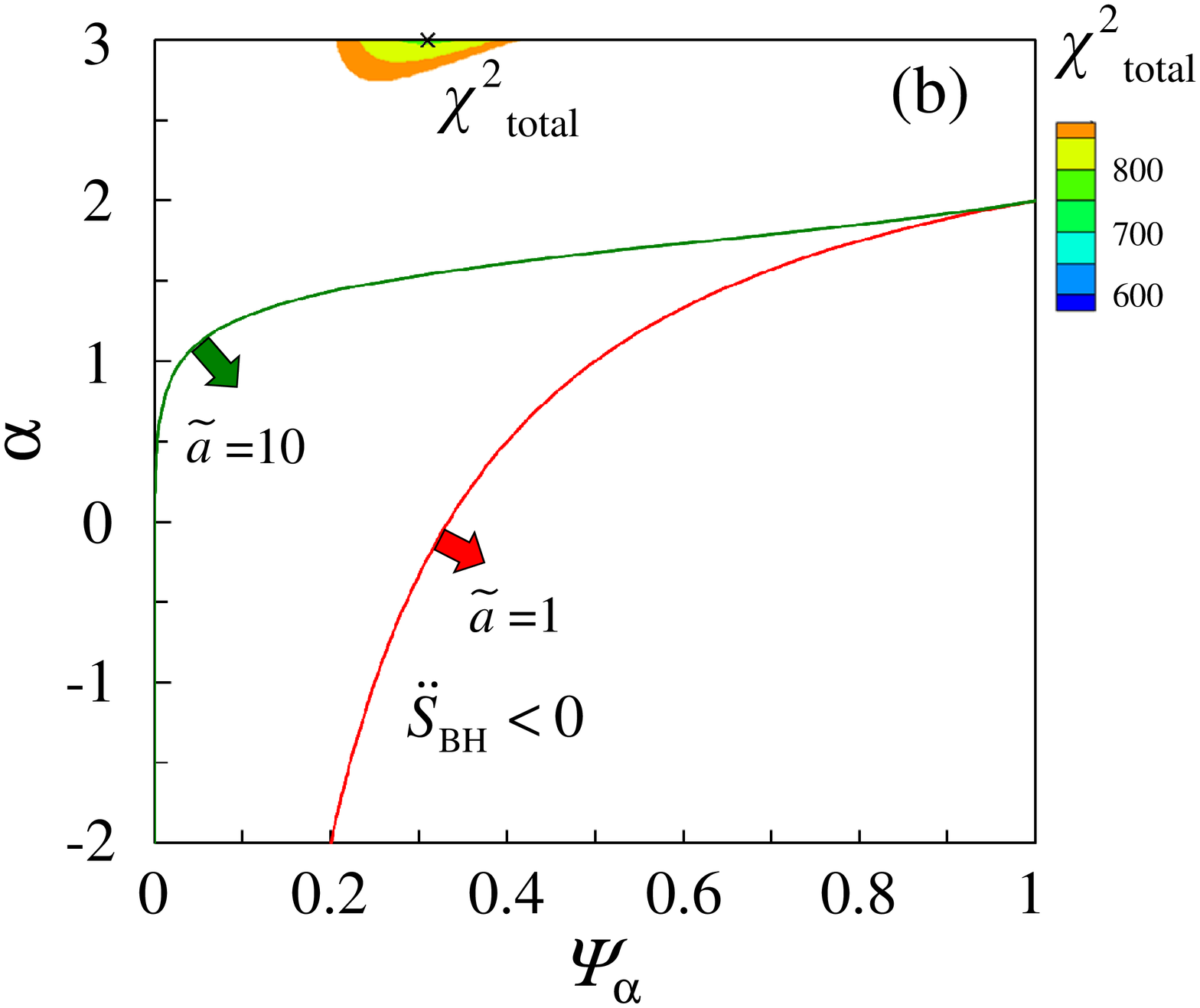}}\\
  \end{center}
 \end{minipage}
\caption{ (Color online). 
Contours of $\chi_{\textrm{total}}^{2}$ in the $(\Psi_{\alpha}, \alpha)$ plane.
(a) $\Lambda (t)$-$H^{\alpha}$ model.  (b)  BV-$H^{\alpha}$ model.
Small-$\chi_{\textrm{total}}^{2}$ regions $(\chi_{\textrm{total}}^{2} < 900)$ are displayed.
The boundary of $\ddot{S}_{\rm{BH}} = 0$ shown in Fig.\ \ref{Fig-SN_GFfsigma8_dS2BH=0_plane} is also plotted.
For the boundary, see the caption of Fig.\ \ref{Fig-SN_GFfsigma8_dS2BH=0_plane}.
In (a), the point labeled $\Lambda$CDM represents $(\Psi_{\alpha}, \alpha)= (0.685, 0)$, corresponding to that model.
The x's mark the locations of the minimum value of $\chi_{\textrm{total}}^{2}$.
The minimum values are summarized in Table\ \ref{tab-chi2min}.
}
\label{Fig-chi2total_dS2BH=0_plane}
\end{figure*}
\begin{table*}[tb]
\caption{Minimum values of chi-squared functions for the $\Lambda (t)$-$H^{\alpha}$ and BV-$H^{\alpha}$ models. 
The location for the minimum value, $(\Psi_{\alpha}, \alpha)$, is also shown.
$\Psi_{\alpha}$ and $\alpha$ are sampled in steps of $0.005$ and $0.025$, respectively.
The locations for $\chi_{\textrm{SN},min}^{2}$ and $\chi_{\textrm{GR},min}^{2}$ are plotted in Fig.\ \ref{Fig-SN_GFfsigma8_dS2BH=0_plane} and the location for $\chi_{\textrm{total},min}^{2}$ is plotted in Fig.\ \ref{Fig-chi2total_dS2BH=0_plane}.
 }
\label{tab-chi2min}
\newcommand{\m}{\hphantom{$-$}}
\newcommand{\cc}[1]{\multicolumn{1}{c}{#1}}
\renewcommand{\tabcolsep}{1.0pc} 
\renewcommand{\arraystretch}{1.25} 
\begin{tabular}{@{}l|ll|ll|ll}
\hline
\hline
$\textrm{Model}$                     &  $\chi_{\textrm{SN},min}^{2}$   &  $(\Psi_{\alpha}, \alpha)$   &   $\chi_{\textrm{GR},min}^{2}$   &  $(\Psi_{\alpha}, \alpha)$        &   $\chi_{\textrm{total},min}^{2}$   &  $(\Psi_{\alpha}, \alpha)$  \\
\hline
$\Lambda (t)$-$H^{\alpha}$     &  $597.2$                                &  $(0.415, 3.000)$                &   $11.5$                                  &  $(0.775, 0.000)$                    &   $630.2$                                   & $(0.580, 0.250)$                \\
BV-$H^{\alpha}$                      &  $597.2$                                &  $(0.415, 3.000)$                &   $11.7$                                  &  $(0.160, 0.900)$                    &   $786.8$                                   &  $(0.310, 3.000)$               \\
\hline
\hline
\end{tabular}\\
\end{table*}

So far, we have considered the specific case of $\Psi_{\alpha}= 0.685$.
In this subsection, we examine constraints on the $\Lambda (t)$-$H^{\alpha}$ and BV-$H^{\alpha}$ models in the $(\Psi_{\alpha}, \alpha)$ plane and discuss the question, "Which model is favored?"
To this end, we provide an overview of the observational and thermodynamic constraints on the two models.
In this analysis, $H_{0}$ is set to $67.4$ km/s/Mpc from the Planck2018 results \cite{Planck2018}.

To examine observational constraints on the two models, we perform a chi-squared analysis using a distance modulus $\mu$ and a combination value $f(z) \sigma_{8}(z)$.
The distance modulus $\mu$ is defined as
\begin{equation}
 \mu = 5 \log d_{L} + 25   ,
\end{equation}
where the luminosity distance $d_{L}$ is given by Eq.\ (\ref{eq:dL}).
The observed distance modulus $\mu$ is obtained from the supernova data \cite{Suzuki_2012}.
The chi-squared function $\chi_{\textrm{SN}}^{2}$ for the supernovae is given by
\begin{equation}
\chi_{\textrm{SN}}^{2} (\Psi_{\alpha}, \alpha)
= \sum\limits_{i=1}^{580} { \left[ \frac{   
                                                         \mu_{\textrm{obs}} (z_{i} ) - \mu_{\textrm{cal}} (z_{i}, \Psi_{\alpha}, \alpha )  }{ \sigma_{i}^{\textrm{SN}} }   
                                     \right]^{2}   }   ,
\label{eq:chi_SN}
\end{equation}
where $\mu_{\textrm{obs}} (z_{i} )$ and $ \mu_{\textrm{cal}} (z_{i}, \Psi_{\alpha}, \alpha ) $ are the observed and calculated distance moduli, respectively, and $\sigma_{i}^{\textrm{SN}}$ is the uncertainty in the observed distance modulus.
The \textit{Union 2.1} set of $580$ type Ia supernovae \cite{Suzuki_2012} is used for the observed data points (numbered $i=1$ to $580$), which are shown in Fig.\ \ref{Fig-dL-z_power}.
Using a combination value $f(z) \sigma_{8}(z)$, the chi-squared function $\chi_{\textrm{GR}}^{2}$ for the growth rate is given by
\begin{align}
& \chi_{\textrm{GR}}^{2} (\Psi_{\alpha}, \alpha) = \notag \\
& \sum\limits_{i=1}^{18} { \left[ \frac{   
                                                         f_{\textrm{obs}}(z_{i} ) \sigma_{8}^{\textrm{obs}} (z_{i} ) - f_{\textrm{cal}} (z_{i}, \Psi_{\alpha}, \alpha ) \sigma_{8}^{\textrm{cal}} (z_{i}, \Psi_{\alpha}, \alpha )  }{ \sigma_{8,i}^{\textrm{GR}} }   
                                     \right]^{2}   }   ,
\label{eq:chi_GR_fsigma8}
\end{align}
where $f_{\textrm{obs}}(z_{i} ) \sigma_{8}^{\textrm{obs}} (z_{i} )$ and $  f_{\textrm{cal}} (z_{i}, \Psi_{\alpha}, \alpha ) \sigma_{8}^{\textrm{cal}} (z_{i}, \Psi_{\alpha}, \alpha ) $ are the observed and calculated values, respectively, and $\sigma_{8,i}^{\textrm{GR}}$ is the uncertainty in the observed value.
The observed data points (numbered $i=1$ to $18$) are taken from the summary in Ref.\ \cite{Nesseris2017} and are shown in Fig.\ \ref{Fig-f(z)sigma8-z_power}.
Each original data point is given in Refs.\ \cite{58,59,60,61,62,63,67,68,69,71,72,73,74,75}.

In addition, a joint chi-squared analysis is performed using the two chi-squared functions.
For the joint chi-squared analysis, the combined chi-squared function $ \chi_{\textrm{total}}^{2}$ is defined by
\begin{equation}
 \chi_{\textrm{total}}^{2} =  \chi_{\textrm{SN}}^{2} + \chi_{\textrm{GR}}^{2}   .
\label{eq:chi_total}
\end{equation}
For these analyses, $\Psi_{\alpha}$ and $\alpha$ are treated as free parameters.
$\Psi_{\alpha}$ is sampled in the range from $0$ to $1$ in steps of $0.005$ and $\alpha$ is sampled in the range from $-2$ to $3$ in steps of $0.025$.
In the present study, $\alpha = 2 - \epsilon$ is sampled instead of $\alpha =2$, to avoid a division by zero, where $\epsilon= 10^{-7}$.
Also, $\Psi_{\alpha}=1$ is not sampled.

We now provide an overview of the observational and thermodynamic constraints on the $\Lambda (t)$-$H^{\alpha}$ and BV-$H^{\alpha}$ models.
Figure\ \ref{Fig-SN_GFfsigma8_dS2BH=0_plane} shows the contours of $\chi_{\textrm{SN}}^{2}$ and $\chi_{\textrm{GR}}^{2}$ in the $(\Psi_{\alpha}, \alpha)$ plane.
In addition, the boundary of $\ddot{S}_{\rm{BH}} = 0$ for $\tilde{a}=1$ and $10$ shown in Fig.\ \ref{Fig-dS2dt2_plane_power} is plotted again in this figure. 
First, we focus on the contours of $\chi_{\textrm{SN}}^{2}$ for the supernovae, which is related to the background evolution of the universe.
Small-$\chi_{\textrm{SN}}^{2}$ regions corresponding to $\chi_{\textrm{SN}}^{2} < 640$ are displayed in Fig.\ \ref{Fig-SN_GFfsigma8_dS2BH=0_plane}. 
The contours of $\chi_{\textrm{SN}}^{2}$ for the two models are the same because the background evolution for both models is equal. 
The region surrounded by the contours indicates a favored region.
For $\chi_{\textrm{SN}}^{2}$, $\Psi_{\alpha} \approx 0.4$--$0.7$ is likely favored.
(We note that $\alpha > 2$ should not satisfy an initially decelerating and then accelerating universe, as examined in Ref.\ \cite{Koma15}.)

Next, we focus on the contours of $\chi_{\textrm{GN}}^{2}$ for the growth rate, which is related to density perturbations.
Small-$\chi_{\textrm{GR}}^{2}$ regions corresponding to $\chi_{\textrm{GR}}^{2} < 80$ are displayed in Fig.\ \ref{Fig-SN_GFfsigma8_dS2BH=0_plane}.
The regions surrounded by contours for the two models are different from each other.
For the $\Lambda (t)$-$H^{\alpha}$ model, $\Psi_{\alpha} \approx 0.1$--$1$ is likely favored [Fig.\ \ref{Fig-SN_GFfsigma8_dS2BH=0_plane}(a)].
In contrast, for the BV-$H^{\alpha}$ model, a low-$\Psi_{\alpha}$ region, specifically $\Psi_{\alpha} \approx 0.1$--$0.2$, is likely favored [Fig.\ \ref{Fig-SN_GFfsigma8_dS2BH=0_plane}(b)].
In particular, for the $\Lambda (t)$-$H^{\alpha}$ model, the regions surrounded by contours of $\chi_{\textrm{GR}}^{2}$ and $\chi_{\textrm{SN}}^{2}$ partially overlap each other [Fig.\ \ref{Fig-SN_GFfsigma8_dS2BH=0_plane}(a)].
However, for the BV-$H^{\alpha}$ model, the contours of $\chi_{\textrm{GR}}^{2}$ and $\chi_{\textrm{SN}}^{2}$ do not overlap [Fig.\ \ref{Fig-SN_GFfsigma8_dS2BH=0_plane}(b)].
Accordingly, the $\Lambda (t)$-$H^{\alpha}$ model is expected to agree more closely with the combined observation data, compared with the BV-$H^{\alpha}$ model. 
In addition, we discuss thermodynamic constraints on the two models.
As shown in Fig.\ \ref{Fig-SN_GFfsigma8_dS2BH=0_plane}(a), for the $\Lambda (t)$-$H^{\alpha}$ model, the above-mentioned overlapped region currently satisfies $\ddot{S}_{\rm{BH}} < 0$, that is, $\tilde{a}=1$.
In contrast, for the BV-$H^{\alpha}$ model, the region surrounded by the $\chi_{\textrm{GN}}^{2}$ contours does not currently satisfy $\ddot{S}_{\rm{BH}} < 0$ [Fig.\ \ref{Fig-SN_GFfsigma8_dS2BH=0_plane}(b)].
These results imply that the $\Lambda (t)$-$H^{\alpha}$ model should agree with observations and satisfy the thermodynamic constraints.
That is, the $\Lambda (t)$-$H^{\alpha}$ model for a non-dissipative universe is expected to be favored.
To confirm this expectation, we examine a combined chi-squared function $\chi_{\textrm{total}}^{2}$, which is calculated from Eq.\ (\ref{eq:chi_total}).

Figure\ \ref{Fig-chi2total_dS2BH=0_plane} shows contours of $\chi_{\textrm{total}}^{2}$ in the $(\Psi_{\alpha}, \alpha)$ plane.
Small-$\chi_{\textrm{total}}^{2}$ regions corresponding to $\chi_{\textrm{total}}^{2} < 900$ are displayed in this figure.
The region surrounded by contours for the $\Lambda (t)$-$H^{\alpha}$ model is wide in comparison with that for the BV-$H^{\alpha}$ model.
In addition, the minimum value of $\chi_{\textrm{total}}^{2}$ for the $\Lambda (t)$-$H^{\alpha}$ model is smaller than that for the BV-$H^{\alpha}$ model.
The minimum value is summarized in Table\ \ref{tab-chi2min}.
(An overview of the constraints is examined here; therefore, the best-fit value is not discussed.)

As shown in Fig.\ \ref{Fig-chi2total_dS2BH=0_plane}(b), the region surrounded by contours for the BV-$H^{\alpha}$ model does not satisfy $\ddot{S}_{\rm{BH}} < 0$ for both $\tilde{a}=1$ and $10$.
In fact, even in the last stage, the region for the BV-$H^{\alpha}$ model does not satisfy $\ddot{S}_{\rm{BH}} < 0$ because the region is outside $\alpha < 2$. 
In contrast, most of the region surrounded by contours for the $\Lambda (t)$-$H^{\alpha}$ model currently satisfies $\ddot{S}_{\rm{BH}} < 0$, especially for small values of $|\alpha|$, as shown in Fig.\ \ref{Fig-chi2total_dS2BH=0_plane}(a).
Accordingly, the $\Lambda (t)$-$H^{\alpha}$ model in this region should be favored over the BV-$H^{\alpha}$ model.
The small $|\alpha|$ region includes a point $(\Psi_{\alpha}, \alpha)= (0.685, 0)$, which corresponds to the standard $\Lambda$CDM model [Fig.\ \ref{Fig-chi2total_dS2BH=0_plane}(a)].
 (The location for $\chi_{\textrm{total},min}^{2}$ slightly deviates from the point.
The small $|\alpha|$ may be related to a weak entanglement of quantum fields between the inside and outside of the horizon.)

\begin{figure} [b] 
\begin{minipage}{0.495\textwidth}
\begin{center}
\scalebox{0.33}{\includegraphics{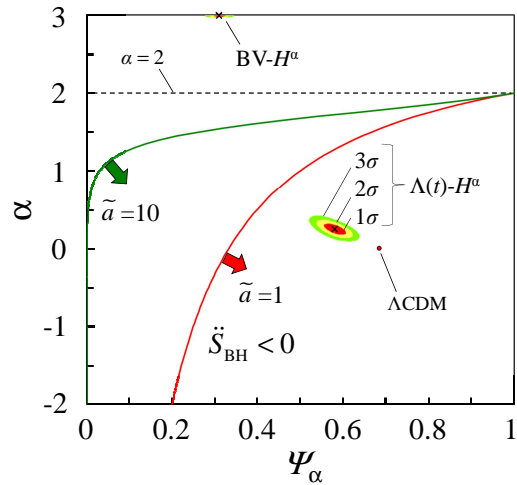}}
\end{center}
\end{minipage}
\caption{ (Color online). Contours of the normalized $L_\textrm{total}$ in the $(\Psi_{\alpha}, \alpha)$ plane  for the $\Lambda (t)$-$H^{\alpha}$ and BV-$H^{\alpha}$ models.
The contours of the $1 \sigma$, $2 \sigma$, and $3 \sigma$ confidence levels are plotted.
The horizontal dashed line represents $\alpha=2$.
The boundary of $\ddot{S}_{\rm{BH}} = 0$, which is shown in Figs.\ \ref{Fig-SN_GFfsigma8_dS2BH=0_plane} and \ref{Fig-chi2total_dS2BH=0_plane}, is also plotted.
For an explanation of this boundary, see the caption of Fig.\ \ref{Fig-SN_GFfsigma8_dS2BH=0_plane}.
The point labeled $\Lambda$CDM represents $(\Psi_{\alpha}, \alpha)= (0.685, 0)$ for that model.
The x within the contours marks the location  for $L_{\textrm{total},max}$, which is equivalent to that for $\chi_{\textrm{total},min}^{2}$ shown in Fig.\ \ref{Fig-chi2total_dS2BH=0_plane} and Table\ \ref{tab-chi2min}.
Note that the region surrounded by contours for the BV-$H^{\alpha}$ model is very small (see the text for further discussion).
 }
\label{Fig-Ltotal_dS2BH=0_plane}
\end{figure}

Finally, a joint likelihood analysis is performed.
For the joint likelihood analysis, a combined likelihood function $L_{\textrm{total}}$ is defined by \cite{Koma8}
\begin{equation}
L_{\textrm{total}} = L_{\textrm{SN}} \times L_{\textrm{GR}} , 
\label{eq:L-joint}
\end{equation}
where $L_{\textrm{SN}}$ and $L_{\textrm{GR}}$ are given by 
\begin{equation}
L_{\textrm{SN}} \propto \exp \left ( {   \frac{- \chi^{2}_{\textrm{SN}} }{2}  } \right)  \quad \textrm{and} \quad  L_{\textrm{GR}} \propto \exp \left ( {   \frac{- \chi^{2}_{\textrm{GR}} }{2}  } \right)  .
\label{eq:L-chi}
\end{equation}
In this study, Eq.\ (\ref{eq:L-joint}) is normalized by the maximum value of $L_{\textrm{total}}$, namely $L_{\textrm{total},max}$.
The normalized $L_{\textrm{total}}$ for the $\Lambda (t)$-$H^{\alpha}$ and BV-$H^{\alpha}$ models can be calculated from the results shown in Figs.\ \ref{Fig-chi2total_dS2BH=0_plane}(a) and (b), respectively.
Using the normalized $L_{\textrm{total}}$, the contours of the $1 \sigma$, $2 \sigma$, and $3 \sigma$ confidence levels are plotted in Fig.\ \ref{Fig-Ltotal_dS2BH=0_plane}.
Here, $1 \sigma$, $2 \sigma$, and $3 \sigma$ correspond to the normalized $ L_\textrm{total} = 3.17 \times 10^{-1}$, $4.60 \times 10^{-2}$, and $2.73 \times 10^{-3}$, respectively \cite{Valent2015,Koma8}.
Accordingly, the region surrounded by the contour for the $3 \sigma$ is narrow, compared with the region for $\chi_{\textrm{total}}^{2} <900$ shown in Fig.\ \ref{Fig-chi2total_dS2BH=0_plane}. 
In particular, the region for the BV-$H^{\alpha}$ model is very small, as shown in Fig.\ \ref{Fig-Ltotal_dS2BH=0_plane}, although the location for $L_{\textrm{total},max}$ is the same as that for $\chi_{\textrm{total},min}^{2}$.
From Fig.\ \ref{Fig-Ltotal_dS2BH=0_plane}, we can confirm that the region surrounded by the contour for the $\Lambda (t)$-$H^{\alpha}$ model currently satisfies $\ddot{S}_{\rm{BH}} < 0$, whereas the region for the BV-$H^{\alpha}$ model does not.

This section examines the observational and thermodynamic constraints on the $\Lambda (t)$-$H^{\alpha}$ and BV-$H^{\alpha}$ models.
Consequently, the $\Lambda (t)$-$H^{\alpha}$ model is found to be favored, compared with the BV-$H^{\alpha}$ model considered here.
In other words, the non-dissipative universe described by the $\Lambda (t)$-$H^{\alpha}$ model is likely consistent with our Universe.
It should be noted that BV-$H^{\alpha}$ models for $\alpha =0$, namely the BV-$H^{0}$ models, agree with the observed supernova and growth-rate data, if negative $c_{\rm{eff}}^{2}$ \cite{Lima2011} and clustered matter \cite{Ramos_2014} can be assumed.
However, these assumptions were not used in this study, and
detailed analyses are left for future research.

\section{Conclusions}
\label{Conclusions}

Cosmological models can be categorized according to how they handle energy dissipation: $\Lambda(t)$ models are used for a non-dissipative universe, and BV models are used for a dissipative universe. To clarify the difference between the two universes, we have examined density perturbations using two types of holographic cosmological models. 
A power-law term proportional to $H^{\alpha}$ is applied to the $\Lambda(t)$ and BV models to systematically examine the two different universes.
In this study, an equivalent background evolution of the universe was set for both the $\Lambda(t)$-$H^{\alpha}$ and BV-$H^{\alpha}$ models.
Based on the background evolution, we derived first-order density perturbations in the two models.
In the derived formulation, $\alpha$ is a free parameter and therefore, the difference in density perturbations between the two models can be systematically examined.

Using the formulation, we examined the evolution of the universe in the $\Lambda(t)$-$H^{\alpha}$ and BV-$H^{\alpha}$ models.
A growth rate $f(z)$ and a combination value $f(z) \sigma_{8}(z)$ for the $\Lambda(t)$-$H^{\alpha}$ model are found to agree with observed data points when $\Psi_{\alpha} =0.685$ (which is equivalent to $\Omega_{\Lambda}$ from the Planck 2018 results), in contrast with the BV-$H^{\alpha}$ model.
In addition, we systematically examined the observational and thermodynamic constraints on the two models by using chi-squared functions in the $(\Psi_{\alpha}, \alpha)$ plane.
Consequently, the $\Lambda(t)$-$H^{\alpha}$ model for small $|\alpha|$ values was found to be consistent with the combined observation data, that is, a distance modulus $\mu$ and $f(z) \sigma_{8}(z)$, and satisfies the maximization of entropy on the horizon of the universe.
This result implies that a $\Lambda(t)$-$H^{\alpha}$ model similar to $\Lambda$CDM models is favored, compared with the BV-$H^{\alpha}$ model examined here.
In other words, the non-dissipative universe described by models like the $\Lambda(t)$-$H^{\alpha}$ model is found to be consistent with our Universe.

Through the present study, we have revealed fundamental properties of the two types of holographic cosmological models in dissipative and non-dissipative universes.
Similar models, including CCDM models for a dissipative universe and $\Lambda(t)$CDM models for a non-dissipative universe, have been separately examined.
The present results should promote the development of a deeper understanding of these cosmological models and bridge the gap between them.

\begin{acknowledgements}
The present study was supported by the Japan Society for the Promotion of Science, KAKENHI Grant Number JP18K03613.
\end{acknowledgements}

\appendix

\section{Unified formulation for the $\Lambda(t)$ and BV models}
\label{Unified formulation}

In Sec.\ \ref{Density perturbations for L(t) and BV}, first-order density perturbations in the $\Lambda(t)$ and BV models are separately presented. 
In this appendix, we review a unified formulation for the $\Lambda (t)$ and BV models using the neo-Newtonian approach proposed by Lima \textit{et al.} \cite{Lima_Newtonian_1997}.
The unified formulation has been examined previously \cite{Koma6} and should be suitable for describing density perturbations in both models systematically.
We review the unified formulation in accordance with the previous work \cite{Koma6}.
In this appendix, general driving terms for the two models, namely $f_{\Lambda}(t)$ and $h_{\textrm{B}}(t)$, are considered without using a power-law term.

In this study, the Friedmann, acceleration, and continuity equations are given by Eqs. (\ref{eq:General_FRW01}), (\ref{eq:General_FRW02}), and (\ref{eq:drho_General}), respectively. In a matter-dominated universe, the continuity equation for a unified formulation can be written as \cite{Koma6}
\begin{equation}
      \dot{\rho} + 3  H   \rho   =   U \rho  , 
\label{eq:drho_U_(p=0)}
\end{equation}
where $U$ is given by
\begin{equation} 
 U = 
    \begin{cases} 
         Q            &   (\Lambda(t)    \hspace{1mm} \rm{model}) ,         \\
         \Gamma  &   (\rm{BV} \hspace{1mm} \rm{model}) .                 \\
    \end{cases}
\label{eq:U_Q_Gamma}
\end{equation}
$Q$ [Eq.\ (\ref{eq:Q_L(t)})] and $\Gamma$ [Eq.\ (\ref{eq:gamma_BV})] are written as
\begin{equation}
  Q =  - \frac{3}{8 \pi G}  \frac{ \dot{f}_{\Lambda}(t)    }{  \rho  }  , 
\label{eq:Q_Unif}
\end{equation}
\begin{equation}
\Gamma = \frac{3}{4 \pi G}  \frac{H h_{\textrm{B}}(t)}{ \rho }      .
\label{eq:gamma_Unif_BV}
\end{equation}
Basic hydrodynamical equations for the neo-Newtonian approach are given in Refs.\ \cite{Lima_Newtonian_1997,Lima2011}.
In fact, for the BV model, the hydrodynamical equations are used in Sec.\ \ref{Formulations for the BV model}.
For the $\Lambda (t)$ model, we use the fundamental equations examined by Arcuri and Waga \cite{Waga1994}.
Consequently, the basic hydrodynamical equations for the unified formulation can be written as \cite{Koma6}
\begin{equation}
 \left ( \frac{ \partial \mathbf{u} }{ \partial t } \right )_{r} + ( \mathbf{u} \cdot  \nabla_{r} ) \mathbf{u}   =  -  \nabla_{r} \Phi - \frac{ \nabla_{r} p_{c} } { \rho + \frac{ p_{c} }{ c^{2} } }   , 
\label{eq:Newtonian_1}
\end{equation}
\begin{equation}
 \left ( \frac{ \partial \rho }{ \partial t } \right )_{r} +  \nabla_{r}  \cdot ( \rho \mathbf{u} ) +  \Theta =  0  , 
\label{eq:Newtonian_2_a}
\end{equation}
\begin{equation}
\nabla_{r} ^{2} \Phi = 4 \pi G  \left ( \rho   +  l \right )   , 
\label{eq:Newtonian_3_a}
\end{equation}
where $\mathbf{u}$ is the velocity of a fluid element of volume and $\Phi$ is the gravitational potential.
For the unified formulation, $\Theta$ and $l$ are given as \cite{Koma6}
\begin{equation} 
 \Theta    =
    \begin{cases} 
         -  Q  \rho        =       \frac{ 3  \dot{f}_{\Lambda}(t)    }{ 8 \pi G }         &   (\Lambda(t)    \hspace{1mm} \rm{model})  ,         \\
         \frac{ p_{c} }{ c^{2} }  \nabla_{r} \cdot \mathbf{u}                               &   (\rm{BV} \hspace{1mm} \rm{model})          ,          \\
    \end{cases}
\label{eq:u_Unif_Theta}
\end{equation}
\begin{equation} 
 l    =
    \begin{cases} 
      -  \frac{ 3 {f}_{\Lambda}(t) }{ 4 \pi G }               &   (\Lambda(t)    \hspace{1mm} \rm{model})  ,         \\
        \frac{ 3 p_{c} }{ c^{2} }                                    &   (\rm{BV} \hspace{1mm} \rm{model})          .         \\
    \end{cases}
\label{eq:u_Unif_l}
\end{equation}
Equations (\ref{eq:Newtonian_1}), (\ref{eq:Newtonian_2_a}), and (\ref{eq:Newtonian_3_a}) are the Euler, continuity, and Poisson equations, respectively. 
Using the basic hydrodynamical equations, we can calculate the time evolution equation for the matter density contrast $\delta$.
Setting $c=1$, using the linear approximation, and neglecting extra terms, we can write the time evolution equation for $\delta$ as \cite{Koma6}
\begin{align}
\ddot{\delta}  & + \left [ H (2  +  3 c_{\rm{eff}}^{2} - 3 u )   - \frac{ \dot{w_{c}} }{ 1+w_{c} }  \right ] \dot{\delta}        \notag \\
                     & +  \Bigg \{      3 (\dot{H} + 2 H^{2}) \left (   c_{\rm{eff}}^{2}  - u \right )     \notag \\
                     & + 3 H  \left [ \dot{c}_{\rm{eff}}^{2}  -  \dot{u}  -  \frac{ \dot{w}_{c} }{ 1+w_{c}  } ( c_{\rm{eff}}^{2} - u )   \right ] \notag \\     
                     &  - 4 \pi G \rho \left ( 1 + w_{c} \right ) ( 1 + 3 c_{\rm{eff}}^{2} )  + \frac{  k^{2} c_{\rm{eff}}^{2} }{ a^{2} }    \Bigg \}     \delta = 0              , 
\label{eq:delta-t_Unif}
\end{align}
where $u$, $w_{c}$, and $c_{\rm{eff}}^{2}$ are defined by 
\begin{equation} 
 u    \equiv  - \frac{U}{3H}      =
    \begin{cases} 
         - \frac{Q}{3H}                                &   (\Lambda(t)    \hspace{1mm} \rm{model})  ,         \\
         - \frac{\Gamma}{3H}  (= w_{c})        &   (\rm{BV} \hspace{1mm} \rm{model})          ,          \\
    \end{cases}
\label{eq:u_Unif_LCDM_BV}
\end{equation}
\begin{equation} 
 w_{c}  \equiv  - \frac{\Gamma}{3H} =
    \begin{cases} 
         0                                       &   (\Lambda(t)    \hspace{1mm} \rm{model})  ,         \\
         - \frac{\Gamma}{3H}           &   (\rm{BV} \hspace{1mm} \rm{model})           ,         \\
    \end{cases}
\label{eq:wc_Unif_LCDM_BV}
\end{equation}
\begin{equation}
 c_{\rm{eff}}^{2}  \equiv   \frac{\delta p_{c} }{\delta \rho }  =
    \begin{cases} 
         0                                                             &   (\Lambda(t)    \hspace{1mm} \rm{model}) ,         \\
         \frac{\delta p_{c} }{\delta \rho }                 &   (\rm{BV} \hspace{1mm} \rm{model}) .                 \\
    \end{cases}
\label{eq:ceff_Unif_LCDM_BV}
\end{equation}
In Eq.\ (\ref{eq:delta-t_Unif}), $\rho$ represents $\bar{\rho}$, that is, a homogenous and isotropic solution for the unperturbed equations. 
The derivation of the above equation is essentially the same as that shown by Jesus \textit{et al.} \cite{Lima2011}.
Note that they assumed $c_{\rm{eff}}^{2} = c_{\rm{eff}}^{2}(t)$ and that the spatial dependence of $\delta$ is proportional to $e^{i \bf{k} \cdot \bf{x} }$, 
where the comoving coordinates $\bf{x}$ are given by $\bf{x} = \bf{r}$$/a$ using the proper coordinates $\bf{r}$ \cite{Lima2011}.

Equation\ (\ref{eq:delta-t_Unif}) is the unified equation for the $\Lambda(t)$ and BV models.
This equation reduces to Eq.\ (\ref{eq:density_L(t)}) in the $\Lambda(t)$ model.
In the BV model, Eq.\ (\ref{eq:delta-t_Unif}) reduces to Eq.\ (\ref{eq:delta-t_BV}), as examined in Ref.\ \cite{Koma6}.

\end{document}